\documentclass[prb, preprint, floatfix, showpacs, amsfonts, amssymb, amsmath]{revtex4-1}
\usepackage{graphicx, times, epsfig, color, float, xcolor, multirow}
\usepackage[latin1]{inputenc}
\usepackage{txfonts}
\def\be{\begin{equation}}
\def\ee{\end{equation}}
\def\bea{\begin{eqnarray}}
\def\eea{\end{eqnarray}}
\def\etal{{\it et al.}}
\def\deg{$^{\circ}$}
\def\asih{{\it a}-Si:H}
\def\asi{{\it a}-Si}
\def\csi{{\it c}-Si}

\definecolor{green1}{RGB}{11,100,35}
\definecolor{maroon}{RGB}{128, 0, 0}
\definecolor{yellow1}{RGB}{254, 223, 0}

\begin{document} 

\title{
Temperature-induced nanostructural evolution of 
hydrogen-rich voids in amorphous silicon: A 
first-principles study 
}

\author{Parthapratim Biswas}
\email[Corresponding author:\,]{partha.biswas@usm.edu}
\affiliation{
Department of Physics and Astronomy, The University
of Southern Mississippi, Hattiesburg, MS 39406, USA}

\author{Durga Paudel}
\email{durga.paudel@usm.edu}
\affiliation{
Department of Physics and Astronomy, The University 
of Southern Mississippi, Hattiesburg, MS 39406, USA}

\author{Raymond Atta-Fynn}
\email{r.attafynn@gmail.com}
\affiliation{Department of Physics, University of Texas
at Arlington, Arlington, TX 76019, USA}

\author{Stephen R. Elliott}
\email{sre1@cam.ac.uk}
\affiliation{Department of Chemistry, University of Cambridge, Cambridge,
CB2 1EW, United Kingdom}

\begin{abstract} 
The paper presents an {\it ab initio} study of temperature-induced 
nanostructural evolution of hydrogen-rich voids in amorphous 
silicon. By using ultra-large {\it a}-Si models, obtained from 
classical molecular-dynamics simulations, with a realistic 
void-volume density of 0.2\%, the dynamics of Si and H atoms on 
the surface of the nanometer-size 
cavities were studied and their effects on the shape and size of the 
voids were examined using first-principles density-functional 
simulations. The results from {\it ab initio} calculations were 
compared with those obtained from using the modified Stillinger-Weber 
potential. The temperature-induced nanostructural evolution of the 
voids was examined by analyzing the three-dimensional distribution 
of Si and H atoms on/near void surfaces using the convex-hull 
approximation, and computing the radius of gyration of the 
corresponding convex hulls. A comparison of the results with 
those from the simulated values of the intensity in small-angle 
X-ray scattering of {\it a}-Si/{\it a}-Si:H  in the Guinier approximation 
is also provided, along with a discussion on the dynamics of 
bonded and non-bonded hydrogen in the vicinity of voids.  
\end{abstract}
\maketitle

\section{Introduction}
Amorphous silicon and its hydrogenated counterpart have a 
wide range of applications, from photovoltaics to thin-film 
technology.~\cite{PV1987,Street2000} Thin films of hydrogenated amorphous silicon 
({\asih}) are extensively used for passivation of crystalline 
silicon (\csi) surfaces, which is essential to 
produce a high open-circuit voltage in silicon-based heterojunction 
solar cells.~\cite{HJT2012,HIT2009}
The passivation of Si dangling bonds by H atoms in the vicinity 
of {\asih}/{\csi} interfaces plays a key role in improving 
the properties of silicon heterostructures.~\cite{Zeman2012}
Post-deposition annealing is routinely used for structural 
relaxation of {\asih} samples and incorporation of hydrogen 
at the {\asih/\csi} interfaces. 
The presence of hydrogen to an extent thus is often 
preferred~\cite{GE2012107} for the preparation of {\asih} 
samples with good interface properties.  Experimental 
studies, such as spectroscopic ellipsometry \cite{SpEll2001,SpEll2003} 
and Raman spectroscopy~\cite{Ramanjap}, indicate that, although 
the presence of nanometer-size voids can degrade the electronic 
quality of the samples, a small concentration of voids 
can improve the degree of surface passivation.  This is 
generally believed to be due to the presence of mobile H atoms 
in a void-rich environment, where 
hydrogen can reach {\asih}/{\csi} interfaces, via diffusion 
or other mechanisms, in comparison to dense {\asih} samples. 

Heat treatment or annealing is an effective tool for 
nanostructural relaxation of laboratory-grown samples. 
Annealing at low to medium temperature (400--700 K) 
can considerably improve the quality of as-deposited 
samples, by removing unwanted impurities and reducing 
imperfections (e.g., defects) in the samples.~\cite{Macco2017}  
For amorphous materials, annealing also increases 
the degree of local ordering in the amorphous environment. 
Earlier studies by small-angle X-ray scattering (SAXS) 
reported the presence of columnar-like geometric 
structure~\cite{NREL_report1994} and blister 
formation in annealed samples of {\asih}.~\cite{Blister2013} 
Likewise, experiments on {\asi}/{\asih}, using positron-annihilation 
lifetime spectroscopy, have indicated that 
annealing above 673 K can lead to the formation of voids 
via vacancy clustering, which affects the nanostructural 
properties of the network.~\cite{Sekimoto2016, Britton2001} It has 
been suggested that at higher temperature, near 800 K, 
a considerable restructuring can take place in the 
vicinity of the void boundary that can modify the 
shape and size of the voids.~\cite{Young2007}  Although tilting 
SAXS~\cite{Young2007,Williamson1989} 
can provide some information on the geometrical shape 
of the voids, a direct experimental determination of 
annealing effects on the morphology of voids in the 
amorphous environment is highly nontrivial, as scattering 
experiments generally provide integrated or scalar 
information on the size and shape of the voids. 
By contrast, computational modeling can yield reliable 
structural information on the morphology of 
voids,~\cite{PBiswas2015,PBiswas2017} provided that 
high-quality ultra-large models of {\asi}/{\asih} -- with 
a linear dimension of several nanometers -- are readily 
available and accurate total-energy functionals to 
describe the atomic dynamics of Si and H atoms are 
in place. However, despite numerous studies on computational 
modeling of {\asi} and {\asih} over the past several decades, 
there exist only a few computational 
studies~\cite{RBiswas_1989, Brahim_2013,BiswasJAP2014,PBiswas2015,PBiswas2017} 
that truly attempted to address the structure and morphology of 
voids on the nanometer length scale in recent years. 

The purpose of the present study is to conduct {\it ab initio} 
simulations of hydrogen-rich voids in ultra-large models 
of {\asi} at low and high annealing temperatures. We are particularly 
interested in the structural reconstruction of the void surfaces, 
induced by annealing, and the resulting effects on the intensity in
small-angle X-ray scattering (SAXS). Further, we intend to study 
the evolution of the shape of voids with temperature in the 
presence of hydrogen inside the voids.
To this end, we develop significantly large 
models of {\asi}, which are characterized by realistic distributions of voids as far as 
the results from SAXS,~\cite{Muramatsu1992,Williamson1989} 
nuclear magnetic resonance (NMR),~\cite{Boyce1985} infrared 
(IR),~\cite{Chabal1984,IRSPec1997} and positron-annihilation~\cite{He1986} 
measurements are concerned.  The term realistic
here refers to the fact that the models must be able to 
accommodate nanometer-size voids with a void-volume fraction 
of 0.2--0.3\%, as observed in SAXS, IR, and NMR measurements.
The three-dimensional shape of the voids are studied by using 
the convex-hull approximation~\cite{PBiswas2015} of the boundary 
region of the voids, upon annealing at 300 K and 800 K, and the 
effects of the shape and size of the voids on the simulated 
SAXS intensity from the models are examined. 
The presence of molecular hydrogen and the 
statistics of various silicon-hydrogen bonding configurations near 
the void surfaces are also examined by introducing H atoms 
within the voids, with a number density of hydrogen (inside 
cavities) as observed in infrared spectroscopy and NMR 
measurements.~\cite{Graebner1984,Chabal1984}  

The rest of the paper is as follows. In sec.\,II, we briefly 
describe the simulation of ultra-large models of {\asi} with a 
linear dimension of a few tens of nanometers and the generation 
of nanometer-size voids in the resulting models 
in order to incorporate a realistic distribution of voids, 
with a void-volume fraction as observed in SAXS and IR measurements.  
Section III discusses the results of our work, with particular 
emphasis on the three-dimensional shape and size of the voids 
upon annealing at 300 K and 800 K. The correlation between 
the three-dimensional shape and size of the voids and the 
intensity of SAXS from simulations is discussed here. This is 
followed by a discussion on the motion of hydrogen in the 
vicinity of the voids, using {\it ab initio} molecular-dynamics 
(AIMD) simulations. This is followed by the conclusions 
in sec.\,IV.  Finally, a comparison of the results with 
those from positron-annihilation lifetime (PAL) 
spectroscopy~\cite{Sekimoto2016} 
and Doppler-broadening positron-annihilation spectroscopy 
(DB-PAS)~\cite{Melskens2017} are presented. 

\section{Computational Methods}
\subsection{Generation of ultra-large models of {\asi}}

We began by generating an ultra-large model of amorphous 
silicon in order to be able to study the intensity due to 
SAXS in the small-angle region of scattering wavevector 
$k \le 1.0$ {\AA}$^{-1}$. 
Since the key purpose of the study is to examine the 
structural evolution of the voids in {\asi} on the nanometer 
length scale, by joint analyses of simulated SAXS 
intensities and three-dimensional distributions of atoms 
near void surfaces,  it is necessary to simulate a sufficiently 
large model that can accommodate a realistic distribution of 
nanometer-size voids with the correct void-volume 
fraction, in the range from 0.01\% to 0.3\%, as observed 
from SAXS and IR measurements.~\cite{Young2007, Williamson1989}  Toward 
this end, a model {\it a}-Si network, consisting 
of 262,400 atoms, was generated using classical 
molecular-dynamics (MD) simulations, followed 
by geometry relaxation, by employing the modified 
Stillinger-Weber potential.~\cite{Vink2001, Stillinger1985}
The initial configuration was obtained by randomly 
generating  atomic positions in a cubic box of length 176.123 {\AA}, which 
corresponds to the experimental mass density, 
2.24 g.cm$^{-3}$, of {\asi}.~\cite{Custer1994}  Thereafter, 
constant-temperature MD simulations were performed 
using a chain of Nos{\'e}-Hoover thermostats~\cite{Nose1984,Hoover1985} 
and a time step of $\Delta t$ = 1 fs.  During the 
MD runs, the system was heated to a temperature of 1800 K for 10 ps, 
and then it was cooled from 1800 K to 300 K at the rate of 
10 K/ps. The heating-cooling cycle was carried out 
for at least 20 iterations. For each cycle, structural properties 
of the system, such as the coordination-number statistics, 
the bond-angle distribution, and the structure factor 
of the evolving configurations were examined.  
The procedure continues until it satisfies the convergence 
criteria that entail imposing a restriction on the upper limit 
of the root-mean-square (RMS) width of the bond-angle 
distribution (${\Delta\theta}_{\text max} \le 10${\deg}) and 
the coordination-defect density (of 3\% or less) in the network. 
It may be noted that these variables are in conflict with 
each other, meaning that one can be reduced at the expense 
of increasing the other, and vice versa. Thus, care was 
taken to ensure that these criteria were simultaneously 
satisfied. 
Once the structural properties satisfied the aforementioned 
convergence criteria, the MD run was terminated and the 
system was relaxed by minimizing the total energy of the 
system with respect to the atomic positions.  The resulting 
configuration was then used as a structural basis for the 
generation of additional models of {\asi} by introducing voids 
in the network. The details of the simulation procedure can 
be found in Ref.\,\onlinecite{JCP2018}.

\subsection{Void geometry and distribution in {\asi}} 
Having obtained an ultra-large model of {\asi}, we proceed
to generate a realistic distribution of voids.  Since 
computational modeling of void formation in {\asi}, by mimicking the actual deposition 
and growth processes, is highly nontrivial, we generated a void 
distribution using experimental information on the shape, size, 
and the number density of voids.  
Experimental data from an array of measurements, such as 
small-angle scattering (SAS),~\cite{DAntonio1979, Muramatsu1992, 
Williamson1989, Wright2007} infrared (IR) spectroscopy,~\cite{Chabal1984}
nuclear magnetic resonance (NMR),~\cite{Boyce1985} positron-annihilation lifetime (PAL) spectroscopy,~\cite{He1986} 
and calorimetry measurements,~\cite{Conradi1981,Graebner1984,Lohneysen1984} 
suggest that the linear size of the voids in {\asi} typically 
lies between 10 {\AA} and 40 {\AA} and that the void-volume fraction 
can range from 0.01\% to 0.3\% of the total volume, depending upon 
the growth process, deposition rate, and the consequent electronic 
quality of the samples. Following these observations, we constructed 
a void distribution in the ultra-large model with the following 
properties: a) the voids are spherical in shape and have a diameter 
of 12 {\AA}; b) the number density of the voids corresponds to a 
void-volume fraction of 0.2\%; c) the voids are sparsely and randomly 
distributed, so that the distance between any two void centers is 
significantly larger than the dimension of the voids.  

\begin{figure}[ht!] 
\centering
\includegraphics[width=0.35\textwidth]{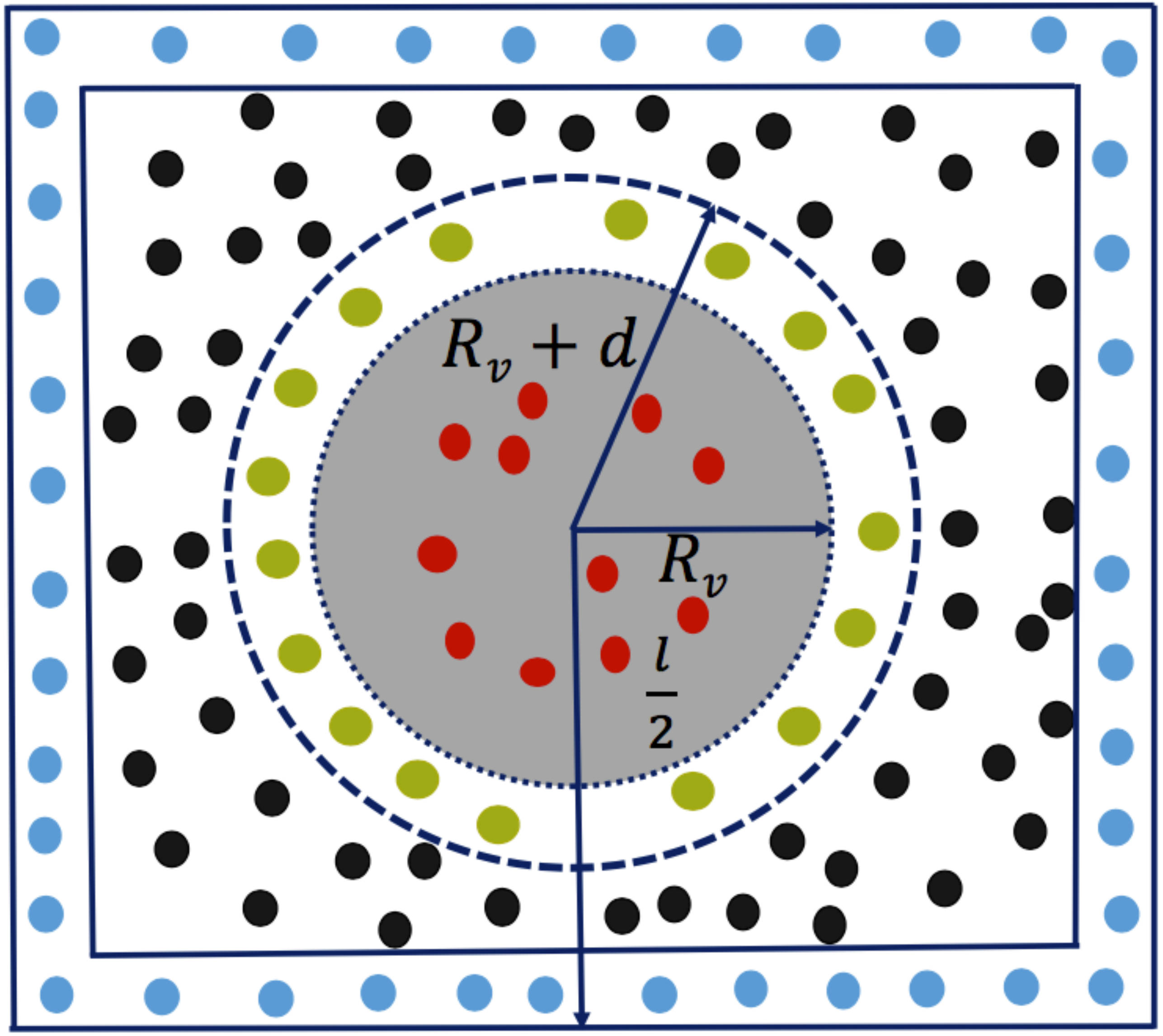}
\caption{
A schematic representation of a void (gray region) in two dimensions. 
The annular region with Si atoms (yellow circles) indicates the void 
boundary, with a few H atoms (red circles) inside the void. 
The bulk Si atoms are shown as black and blue circles -- the latter 
indicate a layer of fixed Si atoms for the purpose of local 
AIMD simulations, as discussed in the text. 
}
\label{void}
\end{figure}

Figure \ref{void} shows an example of a void embedded in a 
two-dimensional amorphous network. 
The void region is indicated by the gray circle 
of radius $R_v$, which is surrounded by a layer of 
Si atoms (yellow) with a thickness $d$.  
Silicon atoms in the layer constitute the 
surface/wall of the void, which we shall refer to 
as the void-surface atoms.  The remaining Si atoms, 
within the cubic box (of length $l$), will be referred to 
as bulk Si atoms and are indicated as black and blue 
circles.  To study the effect of hydrogen on the evolution of void surfaces, 
H atoms were introduced inside the voids so that no two H atoms 
were at a distance less than 1 {\AA} from each other. 
The presence of H atoms in the network requires that 
the problem must be treated at the quantum-mechanical 
level using, for example, first-principles density-functional 
theory. However, given the size of the ultra-large model, 
such a task is hopelessly difficult and computationally 
infeasible, and an approximation of some sort is necessary.  
Here, we employed the local approach to electronic-structure 
calculations by invoking the {\em principle of nearsightedness 
of an equilibrium system}, as proposed by Kohn,~\cite{Kohn1996} 
and address the problem by studying an appropriately 
large sub-system, defined by a cubical box of length $l$ surrounding 
each void (see Fig.\,\ref{void}).  Following an (approximate) 
invariance theorem on electronic structure of disordered 
alloys, due to Heine and coworkers,~\cite{Heine1980,Haydock1975} 
the quantum {\em local} density of states near the 
voids is essentially independent of the boundary conditions 
at a sufficiently large distance,~\cite{boundary} for example, 
the presence of a fixed layer of Si atoms on the edge of the 
box.  Referring to Fig.\,\ref{void}, we chose $R_v$ = 6 {\AA}, 
$d$ = 3 {\AA}, and $l$ = 36 {\AA}. To ensure 
that the void-volume fraction is consistent with the experimentally 
observed value of 0.01\%--0.3\%, we created 12 voids, reflecting a void-volume 
fraction of 0.2\%, which were randomly distributed in the ultra-large model.  
An outer layer of Si atoms (blue) of thickness 3 {\AA} was held 
fixed at the edge of the box of length $l$ for each void during 
first-principles simulations 
and total-energy relaxation of the sub-system so that 
the latter can be put back into the original 
ultra-large model for the calculation of the SAXS 
intensity, upon completion of annealing and geometry 
relaxation without boundary perturbations. 
Each sub-system, consisting of a cubical box of length 36 {\AA} 
with a central void region of radius 6 {\AA}, contained
about 2200 atoms. The size of the cubical box was chosen 
carefully so that the local structural and electronic 
properties of the atoms in the vicinity of void surfaces were 
not affected by a fixed layer of atoms on the edge of the 
box, at a distance of 12 {\AA} from the void surfaces. 
Care was also taken to ensure that the centers of the voids 
were sparsely distributed in the original ultra-large model 
so that the individual boxes or supercells did not overlap 
with each other. To study the effect of hydrogen on void 
surfaces, each void was loaded with 30 H atoms, as observed 
in IR and NMR measurements.\cite{H2_void} 
The initial distributions of the H atoms were generated to 
satisfy the following two criteria: a) the H atoms were 
randomly distributed so that no two H atoms were separated 
by a distance less than 1 {\AA}; b) the distance between 
any H atoms inside a void and a Si atom on the void surface 
was greater than or equal to 2 {\AA}. The latter was achieved 
by confining the H atoms within a sphere of radius 4 {\AA} 
from the center of the void. These criteria prevent the 
system from developing a sudden large force on light H atoms 
and an instantaneous formation of silicon-hydrogen bonds 
on void surfaces at the beginning of {\it ab initio} 
molecular-dynamics (AIMD) simulations. 

\subsection{{\it Ab initio} simulations of voids in {\asi}} 

In the preceding section, we have generated a number of 
models (supercells/subsystems), containing as many as 2200 atoms and a central 
void of radius 6 {\AA}, using the modified 
Stillinger-Weber (SW) potential. However, since the 
SW potential may not  accurately describe the 
chemistry of silicon surfaces, it is necessary to treat these models using 
a first-principles total-energy functional from 
density functional theory (DFT) in order to reduce 
artifacts, such as the number of excessive dangling 
bonds or defects, associated with the SW-generated 
void surfaces.  The standard DFT protocol here is to 
compute the self-consistent-field (SCF) solution of the 
Kohn-Sham (KS) equation, and the presence of voids or 
inhomogeneities in the system suggests that the generalized 
gradient approximation (GGA) should be employed in order 
to deal with the atomic density fluctuations near void 
surfaces. However, it is simply not feasible to solve 
the KS equations self-consistently for a problem of this 
size (i.e., about 2,200 atoms) and complexity using 
the existing plane-wave-based/local-basis DFT 
methodology. In particular, the computational cost 
for conducting SCF AIMD runs for several picoseconds 
is prohibitively large, even for a single configuration. 
To reduce the computational complexity of the problem, 
it is thus necessary to employ a suite of approximations 
for solving the KS equation. Here, we invoke the 
non-self-consistent Harris-functional 
approach~\cite{Harris, Zaremba} in the local density 
approximation (LDA), using a minimal set of basis functions 
for Si atoms.  While these approximations may affect the accurate 
determination of some aspects of the electronic structure 
of void surfaces, the results from test calculations on 1000-atom models with 
a void of radius 5 {\AA}, obtained via the full SCF procedure 
for solving the KS equation in the LDA and GGA, indicate that 
the structural properties of the void surfaces in {\asi} are 
minimally affected by these approximations.
A discussion on the accuracy and computational efficiency 
of the Harris-functional approach to treat {\em bulk} {\asi} 
and {\asih} can be found in Ref.\,\onlinecite{Atta-Fynn2004}. 

{\it Ab initio} calculations and total-energy optimization were 
conducted using the first-principles density-functional code 
{\sc Siesta}.~\cite{siesta1} The latter employs a localized 
basis set and norm-conserving Troullier-Martins 
pseudopotentials,~\cite{TM} which are factorized in the 
Kleinman-Bylander~\cite{KB} form to remove the effect of 
core electrons. Electronic and exchange correlations between 
electrons are handled using the LDA, by employing the 
Perdew-Zunger parameterization of the 
exchange-correlation functional.~\cite{PZ} 
As stated earlier, we employed the Harris-functional 
approach~\cite{Harris,Zaremba} that involves the 
linearization of the Kohn-Sham equations for structural 
relaxation and AIMD simulations. The latter were conducted 
over a period of 10 ps, with a time step of 1 fs.  Single-zeta 
(SZ) and double-zeta-polarized (DZP) basis functions were employed 
for Si and H, respectively. Here, SZ functions 
correspond to one $s$ and three $p$ orbitals for Si atoms, 
whereas DZP functions refer to two $s$ and 
three $p$ orbitals for H atoms. The size of the simulation box 
or supercell is large enough in this work so that the Brillouin 
zone collapses onto the $\Gamma$ point, which was used for the 
Brillouin-zone sampling. 

\subsection{Simulation of SAXS intensity and void-size determination}

\begin{figure}[t]
\centering
\includegraphics[width=0.35\textwidth]{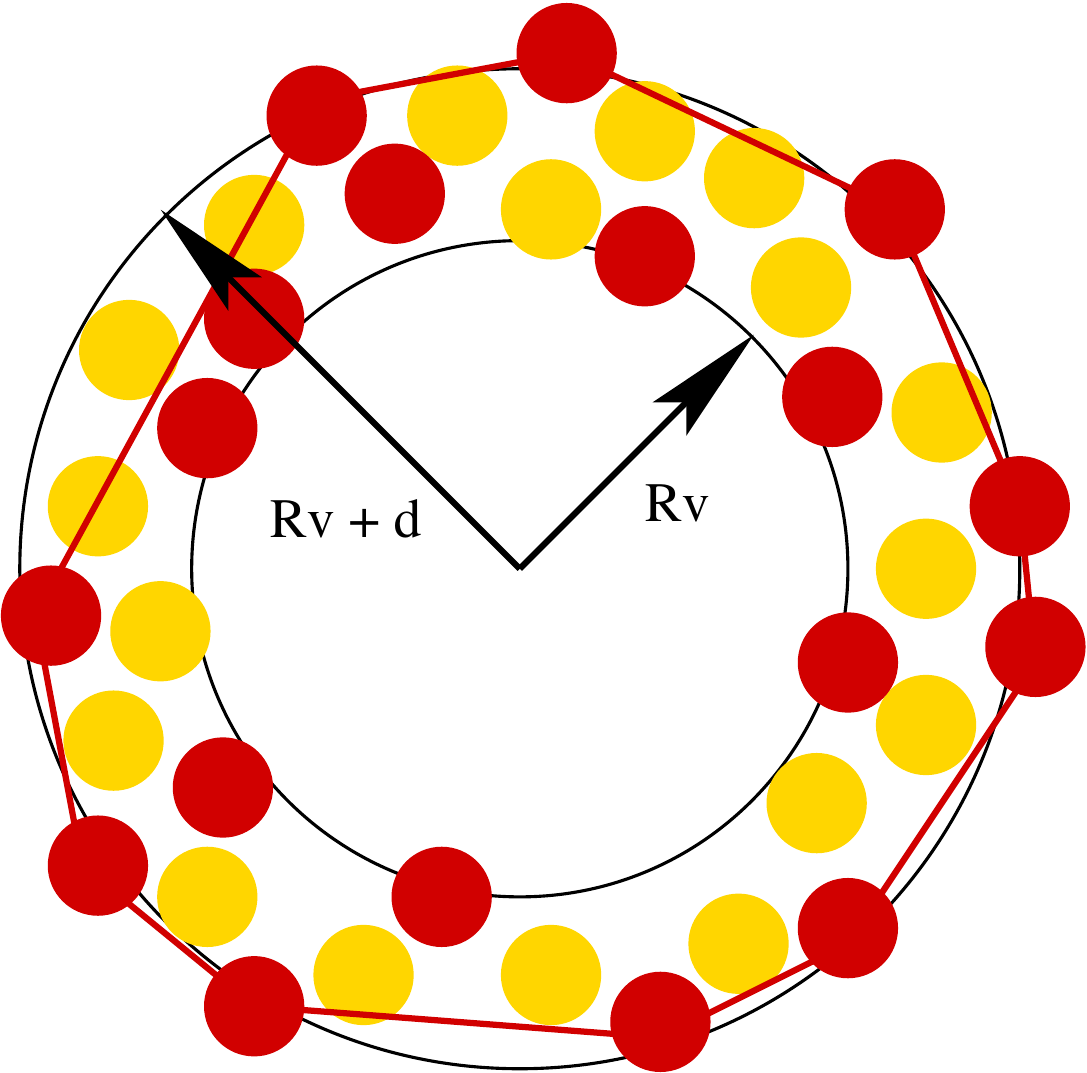}
\caption{
An illustration of the (re)construction of the shape 
and size of a void formed by an assembly of void-surface 
atoms in two dimensions. The distributions of the 
atoms before and after annealing are shown in yellow 
(\textcolor{yellow1}{\large $\bullet$}) and red 
(\textcolor{red}{\large $\bullet$}) colors, 
respectively. The convex polygon (red line) approximates 
the void region in two dimensions after annealing. 
}
\label{FIG2}
\end{figure}

The SAXS intensity for a binary system can be expressed 
with the aid of the partial structure factors and the 
atomic form factors of the constituent atoms. For a binary 
system, the scattering intensity, $I(k)$, can be expressed 
as,~\cite{Cusack_book}
\begin{align}
\frac{I(k)}{N} = & f_{A}^2(x_{A}x_{B} + x_{A}^2 S_{AA}) 
+ f_{B}^2(x_{A}x_{B} + x_{B}^2 S_{BB}) \notag \\ & 
+ 2\, f_{A} f_{B}x_{A}x_{B} (S_{AB} - 1), 
\label{EQ1}
\end{align}
where
\begin{align}
S_{ij}(k) = & 1 + \frac{4\pi n_0}{k}\int _0^{\infty} r\,[g_{ij}(r)-1]\,\sin{kr} \,dr  
\label{EQ2}
\end{align}
is the partial structure factor associated with the atoms of type 
$i$ and $j$ such that $S_{ij} = S_{ji}$, and $g_{ij}(r)$ is the 
partial pair-distribution function.  Here, $x_{i}=\frac {N_i}{N}$ 
and $f_i$ are the atomic fraction and the atomic form factor of 
the atoms of type $i$, respectively. In Eq.\,(\ref{EQ2}), the 
partial pair-distribution function, $g_{ij}(r)$, is defined as 
$g_{ij}(r) = n_{ij}/n_j$, where $n_{ij}$ is the average 
number density of atoms of type $j$, $n_j = x_j n_0$, and 
$n_0 = N/V$. 
Once the scattering intensity is available, either numerically or 
experimentally, the size of the voids, or any extended 
inhomogeneities, can be estimated by invoking the Guinier 
approximation~\cite{Guinier1995} in the small-angle limit, 
$k\,R_G\le1$, 
\begin{equation} 
I(k) = I(k=0)\,\exp\left(-\frac{k^2R_G^2}{3}\right), 
\label{EQ3}
\end{equation}
\noindent 
where $R_G$ is the characteristic size of the inhomogeneities 
in the dilute concentration limit.  While Eq.\,(\ref{EQ3}) provides a reasonable estimate 
of void sizes from experimental data, difficulties arise in 
extracting the $R_G$ value from a $\ln I$-$k^2$ 
plot in simulations, owing to the finite size of the 
models. Since the effective small-angle region of $k$ 
is approximately given by $4\pi/L < k < 1/R_G$, where 
$L$ is the linear dimension of the sample, a sufficiently
large model is needed in order to obtain the mean value 
of $R_G$, via averaging over several independent configurations.  
However, it has been observed~\cite{Paudel2018} that, even for very large 
models, the lower limit of $k$ cannot be reduced arbitrarily 
by increasing $L$ due to the presence of noise in $r\,g(r)$ 
beyond $r \approx $ 25--30 {\AA}.  The term $r\,[g(r)-1]$ 
in Eq.\,(\ref{EQ2}) for $r > $ 25 {\AA} introduces 
artifacts in $S(k)$ for small values of $k$, which make 
it difficult to accurately compute $R_G$ from the simulated 
intensity data.  This necessitates the construction of 
alternative measures of void sizes and shapes directly 
from the distribution of atoms in the vicinity of voids. 
A simple but effective measure is to compute the gyrational 
radius from the distribution of the atoms on void surfaces 
between radii $R_v$ and $R_v+d\,(=R^{\prime}_v)$, where 
$R_v$ is the void radius and $d$ is the width of the 
void surface, as discussed in sec.\,IIB. This is schematically 
illustrated in two dimensions in Fig.\,\ref{FIG2}, where 
the positions of the void-surface atoms before and after 
annealing are shown in yellow and red colors, respectively. 
Assuming that the void-surface atoms (yellow) can displace, 
in the mean-square sense,  from their initial position 
by $x$ during annealing, the radius of gyration, $R_g$, 
for the assembly of void atoms (red) can be expressed as, 
\begin{equation} 
R_g^2=\frac{1}{n_s} \sum_{i=1}^{n_s} (r_i-r_{\rm cm})^2\,H[R^{\prime}_v + x - r_i]. 
\label{4a}
\end{equation}

\noindent 
Here, $r_i$ and $r_{\rm cm}$ are the position of the $i^{th}$ 
atom and the center of mass of $n_s$ void-surface atoms, 
respectively.  The atoms on the reconstructed surface are 
distributed within a spherical region of radius $R^{\prime}_v 
+ x$ and $H[y]$ is the Heaviside step function that asserts 
that only the atoms with $y \ge 0$ contribute to $R_g$. 
\begin{table}[t!]
\caption{\label{TAB1} 
Structural properties of the ultra-large model of {\asi}. 
$N$, $L$, $\rho$, $C_{4}$, $r$, $\theta$, and $\Delta\theta$ 
are the number of atoms, simulation box length, mass density, 
percentage four-fold coordination, average bond length, 
average bond angle (in degree), and the root-mean-square 
width of the bond angles, respectively.
}
\begin{ruledtabular}
\begin{tabular}{cccccccc}
$N$& $L$ ({ \AA})& $\rho$ (g.cm$^{-3}$) & $C_{4} (\%) $&  $r$ ({\AA}) &
$\theta$ &  $\Delta\theta$ \\ \multicolumn{3}{c}{} \\
\hline
$262400$ &$176.123$ &  $2.24$ &  $97.6$ &  $2.39$  &  $109.23$ &  $9.26$      \\ 
\end{tabular}
\end{ruledtabular}
\end{table} 
Here, we chose a value of $x$ = 1.5 {\AA} for Si atoms 
in order to define the boundary of a reconstructed void 
surface.  Since annealing at high temperature can 
introduce considerable restructuring of void surfaces, 
it is often convenient to invoke a suitable convex 
approximation to estimate the size and shape of the 
voids. Here, we employ the convex-hull approximation 
that entails constructing the minimal convex polyhedron that 
includes all the void-surface atoms on the boundary. The size of 
the convex region can be expressed as the radius of gyration 
of the convex polyhedron, or convex hull, 
\begin{equation}
R_H^2=\frac{1}{n_H} \sum_{j=1}^{n_H}(r_j-\bar r_h)^2.
\label{4b}
\end{equation}
In Eq.\,(\ref{4b}), $n_H$ is the number of atoms (or vertices) 
that defines the convex polyhedral surface, and $\bar r_h$ 
is the center of mass of the polyhedron. 
To determine the degree of deviation of the shape of a (convex) 
void from an ideal sphere, we employ the convex-hull 
volume ($V_H$) and surface area ($A_H$) and obtain the sphericity 
parameter,\cite{Wadell1935}
\begin{equation}
\Phi_{S}=\frac {\pi^{\frac{1}{3}} {(6V_{H})}^{\frac{2}{3}}}{A_{H}}. 
\label{EQ6}
\end{equation}
The definition above is frequently used to measure the shape 
and roundness of sedimentary quartz particles, and it expresses 
the ratio of the surface area of a sphere to that of a non-spherical 
particle having an identical volume.~\cite{Wadell1935} 
It may be noted that a highly distorted void may not be 
accurately represented by a convex polyhedron and that 
the convex-hull volume provides an upper bound of the 
actual void volume, leading to $R_H \ge R_g$. 
Likewise, as observed by Porod~\cite{Porod}, the $R_G$ 
value obtained from the Guinier approximation may not 
be accurate for very anisometric voids. 

\begin{figure}[t!] 
\centering
\includegraphics[width=0.5\textwidth]{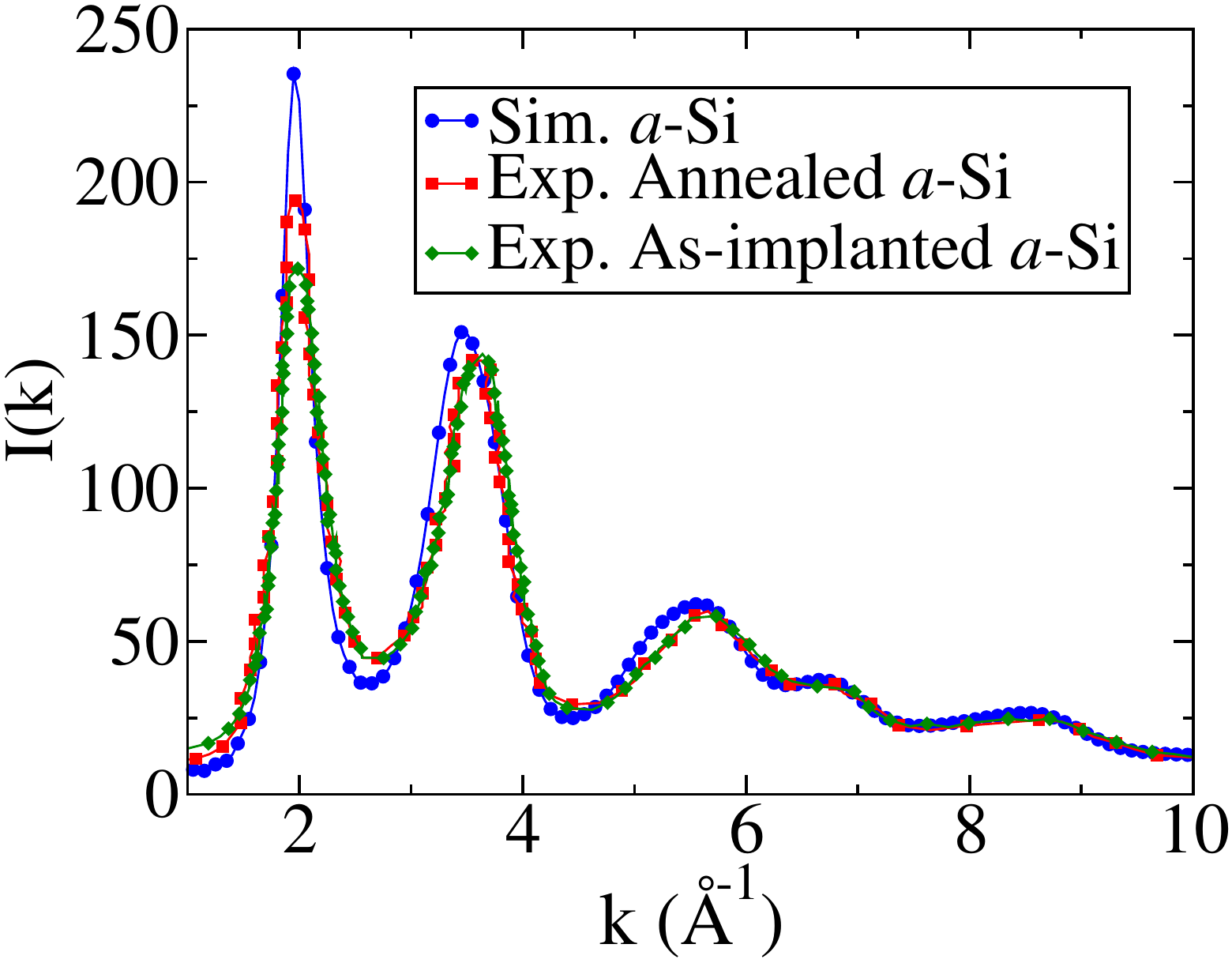}
\caption{
The X-ray scattering intensity for {\asi} from experiments 
and classical molecular-dynamics simulations. The 
experimental data~\cite{Laaziri1999} 
correspond to the results from as-implanted 
(\textcolor{green1}{$\Diamondblack$})
and annealed (\textcolor{red}{\small $\blacksquare$})
samples, while the simulated values 
(\textcolor{blue}{$\bullet$})
of the intensity correspond to the modified Stillinger-Weber 
model of {\asi}.
}
\label{FIG3}
\end{figure}

\section{Results and Discussion}
Before addressing the temperature-induced nanostructural changes 
in voids, we briefly examine the ultra-large 
model of {\asi}, developed in sec.\,IIA, in order to validate its 
structural properties.  Figure \ref{FIG3} shows the 
simulated values of the X-ray scattering intensity, along with the 
experimental values of the intensity for as-implanted 
(green) and annealed (red) samples of pure {\asi}, reported 
by Laaziri {\etal}~\cite{Laaziri1999} The simulated values 
closely match with the experimental data in Fig.\,\ref{FIG3}, 
especially for the annealed samples, in the wide-angle 
region from 1 {\AA}$^{-1}$ to 10 {\AA}$^{-1}$ and beyond. 
Some characteristic structural properties of the ultra-large 
model are also listed in Table \ref{TAB1}. Together with 
the X-ray intensity, the average bond angle and its deviation, 
109.23{\deg}$\pm$ 9.26{\deg}, and the number of four-fold-coordinated 
atoms, 97.6\%, suggest that the structural properties of 
the model obtained from classical molecular-dynamics 
simulations in sec.\,IIA are consistent with 
experimental results.  
It may be reasonably assumed that the presence of a 
small amount of coordination defects ($\approx$ 2.4\%), 
which involve a length scale of 2--3 {\AA} and are 
sparsely distributed in the amorphous environment, 
would not affect the scattering intensity appreciably 
in the small-angle region of the scattering 
wavevector.~\cite{defect} A minor deviation in the 
height of the first peak in the simulated data in 
Fig.\,\ref{FIG3} from experimental values can be 
attributed in part to the use of the classical SW 
potential and partly to the large size of the model. 
A discussion on the validation of the structural 
properties of these models can be found in 
Ref.\,\onlinecite{JCP2018}. 

\begin{figure}[t!]
\centering
\includegraphics[width=0.405\textwidth]{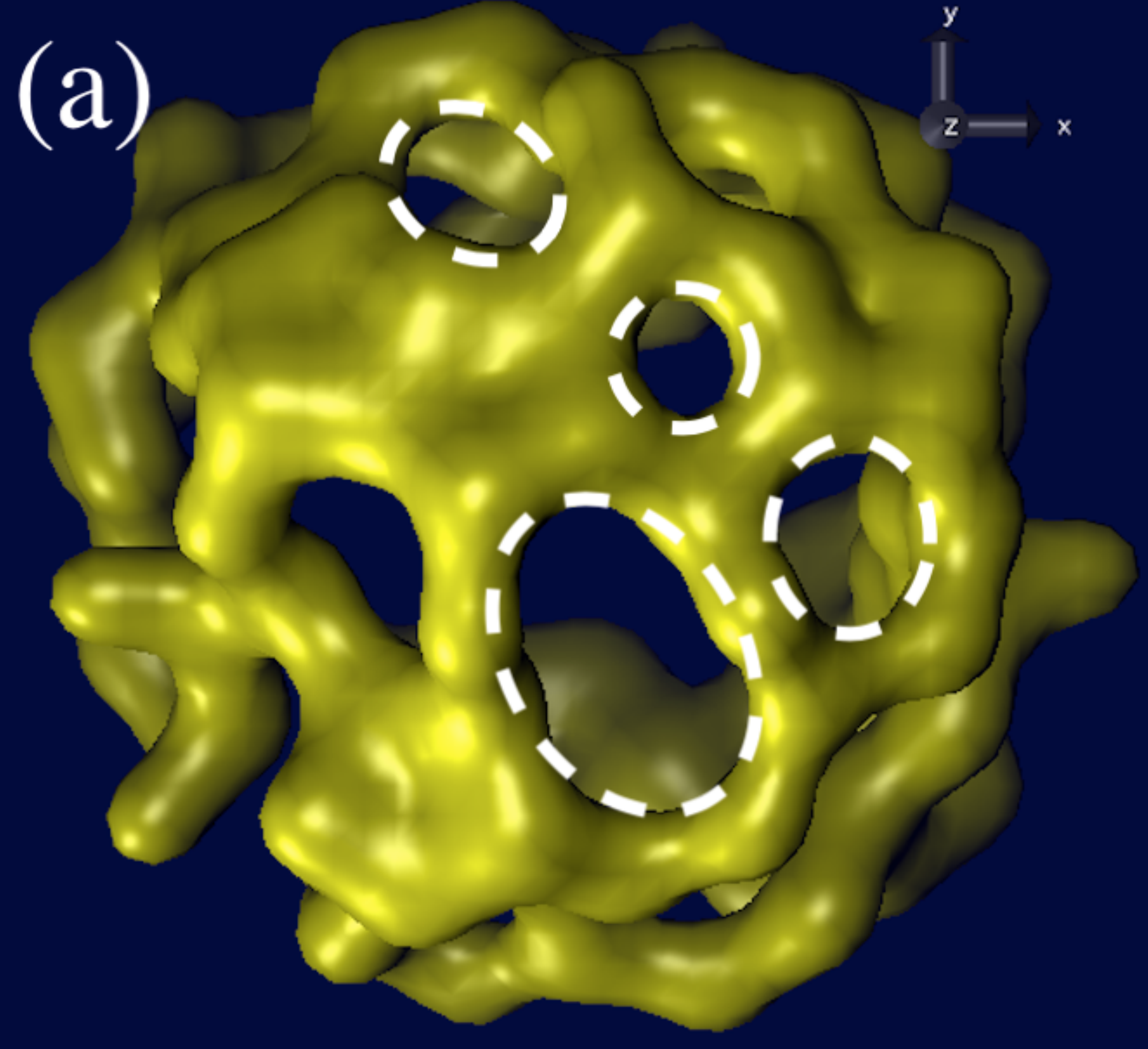}
\includegraphics[width=0.4\textwidth]{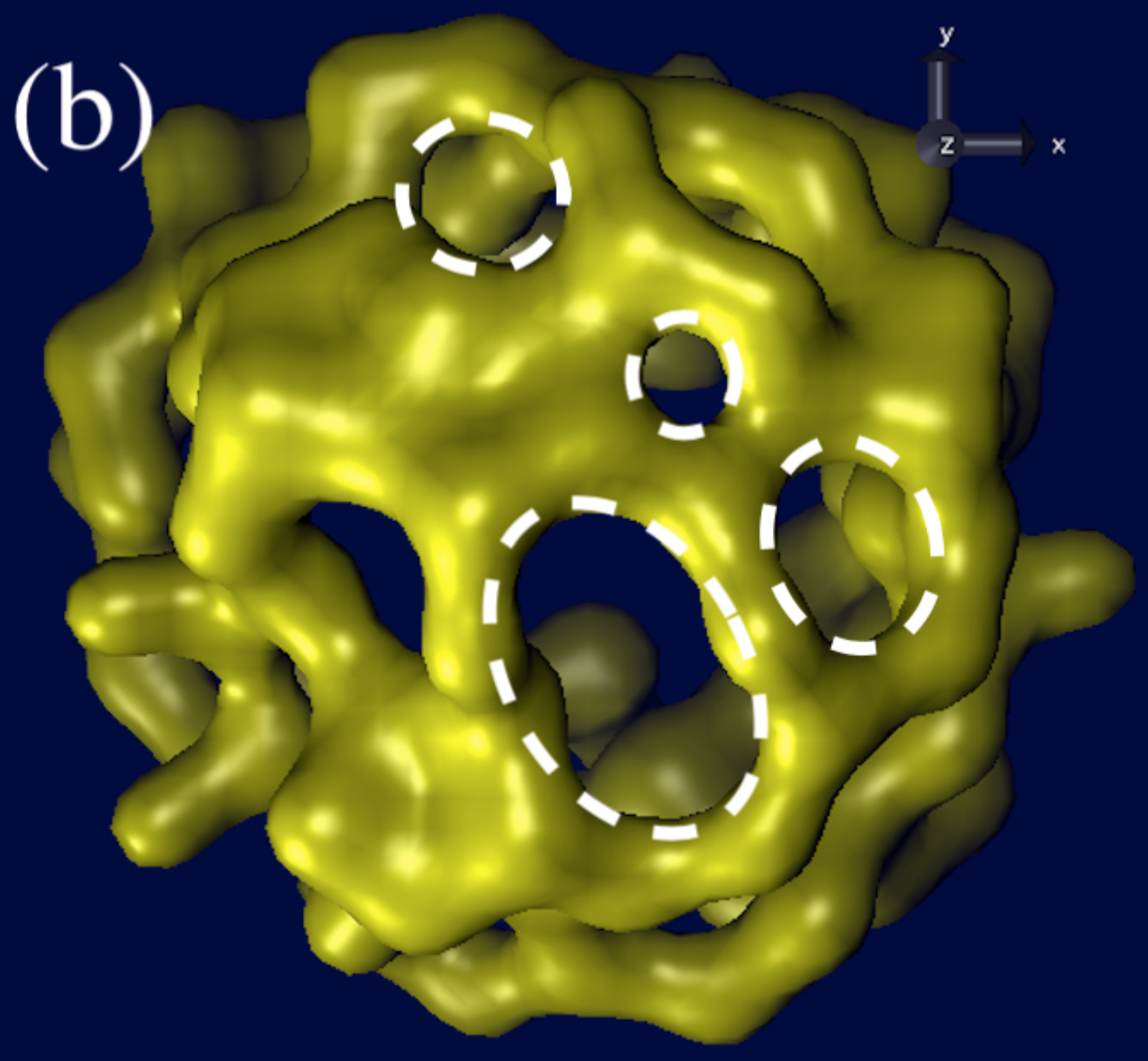} 
\caption{
An isosurface representation of a void (V3), obtained from 
the Gaussian-convolution of atomic positions, after 
thermalization at 300 K from: (a) classical; and (b) 
{\it ab initio} MD simulations. The white rings indicate 
minor structural differences between the two surfaces. 
See sec.\,IIIA for details. 
}
\label{FIG4}
\end{figure}

\subsection{Structural evolution of voids in {\asi} at 300 K and 800 K}

\begin{figure}[t!]
\includegraphics[width=0.5\textwidth]{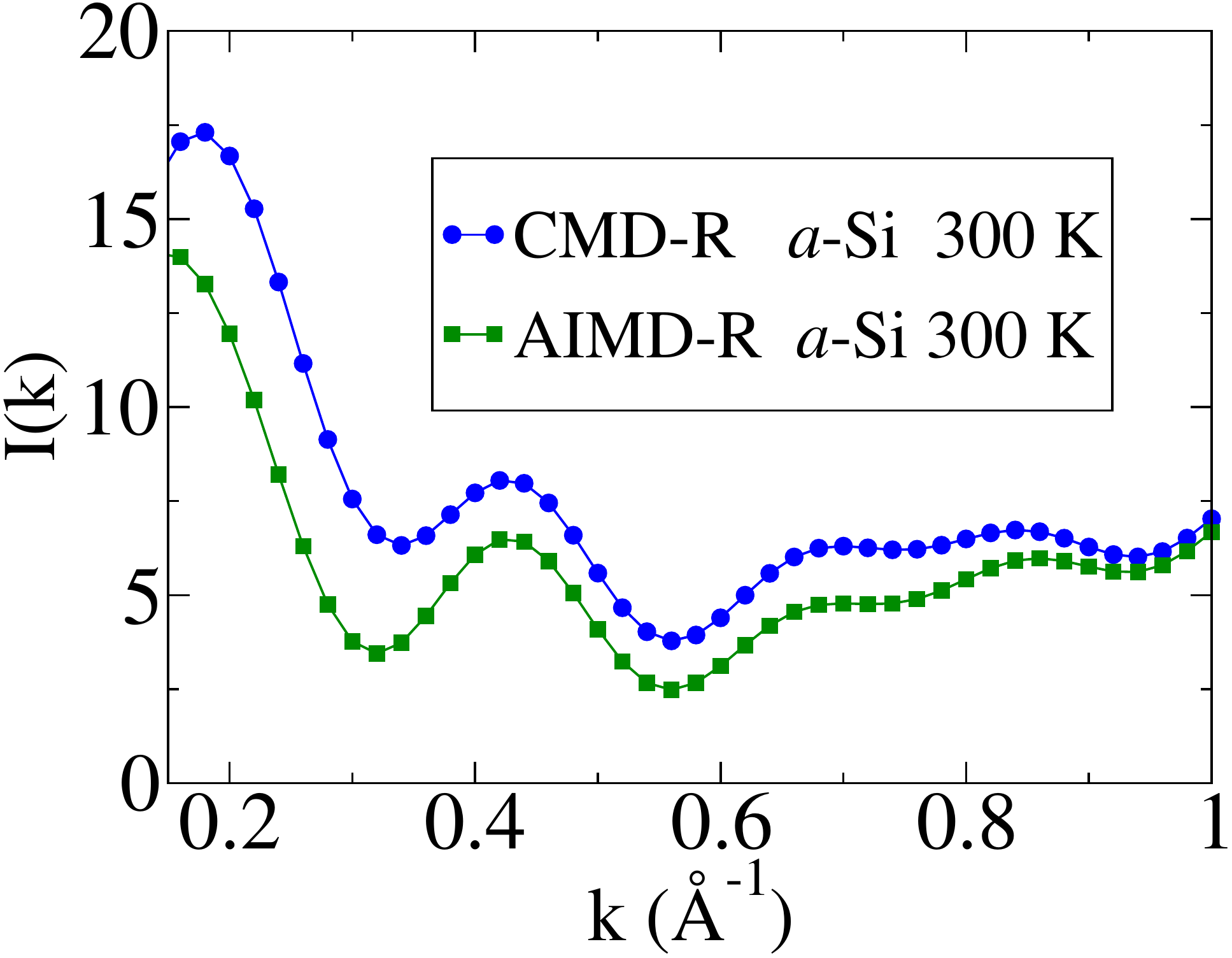}
\caption{
The SAXS intensity for {\asi} from classical (CMD-R) 
and {\it ab initio} (AIMD-R) MD simulations of {\asi} 
at 300 K, followed by geometry relaxation. 
}
\label{FIG5}
\end{figure}

\begin{figure}[t!] 
\includegraphics[width=0.5\textwidth]{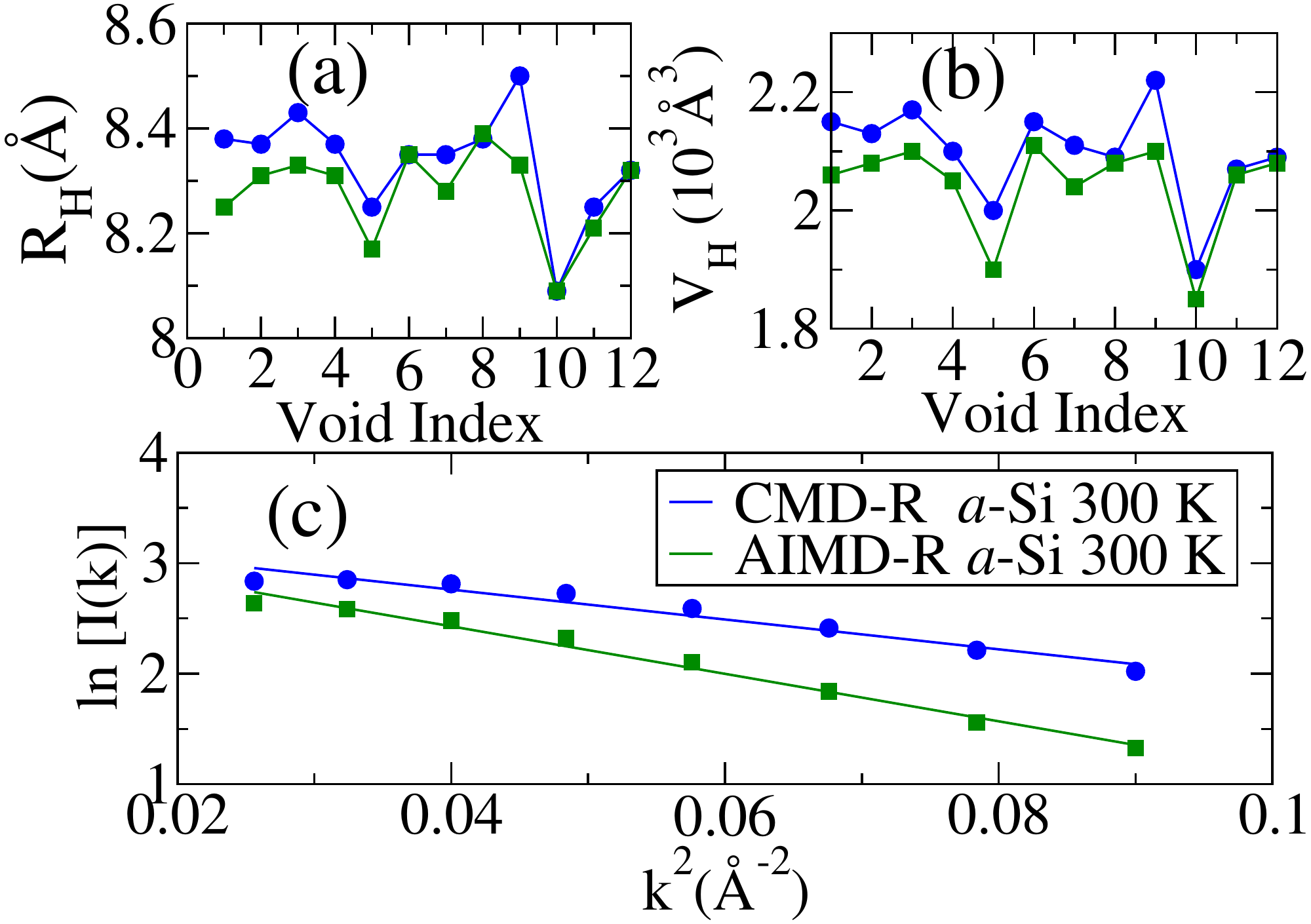}
\caption{
(a) The radius of gyration of the voids from the convex-hull 
approximation at 300 K. The results for the CMD-R 
(\textcolor{blue}{$\bullet$}) 
and AIMD-R  
(\textcolor{green1}{\small $\blacksquare$}) 
models are presented here. 
(b) Estimated void volumes in the convex approximation. 
(c) Guinier plots for the CMD-R and AIMD-R models 
at 300 K. The Guinier radii from the plots correspond 
to a value of 6.38 {\AA} (CMD-R) and 8.07 {\AA} (AIMD-R).
}
\label{FIG6}
\end{figure}

In this section, we discuss the three-dimensional shape and size 
of the voids in pure {\asi} obtained from classical and quantum-mechanical 
annealing of the system at 300 K and 800 K, followed by 
total-energy optimization.  
Starting with an examination of 
the classical- and {\it ab-initio}-generated  void surfaces, 
we analyze the computed SAXS spectrum obtained 
from the ultra-large model, which is embedded with the 
supercells containing voids annealed at 300 K and 800 K.  
Figure \ref{FIG4} shows the structure of a 
representative void (V3) after thermalization 
at 300 K for 10 ps, using classical and {\it ab initio} MD 
simulations. The void surface shown in Fig.\,\ref{FIG4} 
corresponds to a Gaussian-convoluted 
isosurface from the {\sc Xcrysden} package.\cite{xcrysden}
The surface has been generated by placing a three-dimensional Gaussian 
function at the center of each atom and ensuring that the 
function has a value of 2.0 and 1.0 at the center and on the 
surface of the atom, respectively.  By choosing an appropriate 
isovalue, between 0 to 2, and a suitable radius for Si atom, 
from standard crystallographic databases, it is possible to 
examine the morphological changes of voids, 
associated with the structural relaxation of the void-surface 
atoms at different temperature.
A close examination of Figs.\,\ref{FIG4}(a) and 
\ref{FIG4}(b) suggests that the resultant structures are very 
similar to each other as far as the void surfaces are concerned, 
except for small changes as indicated by white rings.  A similar observation 
has been made for the remaining voids as well, which 
showed very little changes on the void surfaces. These results 
are consistent with the variation of the simulated intensity 
shown in Fig.\,\ref{FIG5}. 
A somewhat more pronounced scattering from the classically-treated 
voids (CMD-R) in comparison to the {\it ab-initio}-treated 
voids (AIMD-R) can be attributed to a minor expansion of the 
voids during classical simulations.  This is clearly evident 
from Figs.\,\ref{FIG6}(a) and 
\ref{FIG6}(b), where the radius of gyration and the volume 
of the voids, respectively, are obtained from the convex-hull  
approximation of the void region. 
The polyhedral radii ($R_H$) for the voids from classical 
simulations were found to be consistently larger than 
the corresponding radii from {\it ab initio} simulations 
in Fig.\,\ref{FIG6}(a). A similar conclusion applies to polyhedral 
volumes ($V_H$), which are plotted in Fig.\,\ref{FIG6}(b).  
Once the linear size, area, and volume of the voids are 
available in the convex approximation, the sphericity of 
voids, $\Phi_S$, can be determined from Eq.\,(\ref{EQ6}). 
A further measure of the size of the voids follows from 
the variation of the SAXS intensity, $I(k)$, with the 
wavevector, $k$, in the small-angle region. 
By invoking the Guinier approximation (see Eq.\,\ref{EQ3}), the 
Guinier radius, $R_G$, can be obtained from 
$\ln I(k)$ vs.\;$k^2$ plots.  Figure \ref{FIG6}(c) shows 
the Guinier fits in the wavevector range, 
0.15 {\AA} $\le k\le$ 0.3 {\AA}, which resulted in an $R_G$ value 
of 6.38 {\AA} for the CMD-R model and 8.07 {\AA} for 
the AIMD-R model.  It should suffice to mention that the 
computation of the Guinier radius, $R_G$, is affected by 
the limited data in the small-angle region of the intensity 
spectrum even for our large model and thus it may vary 
from the estimate obtained from the convex approximation.

\begin{figure}[h] 
\centering
\includegraphics[width= 0.4\textwidth]{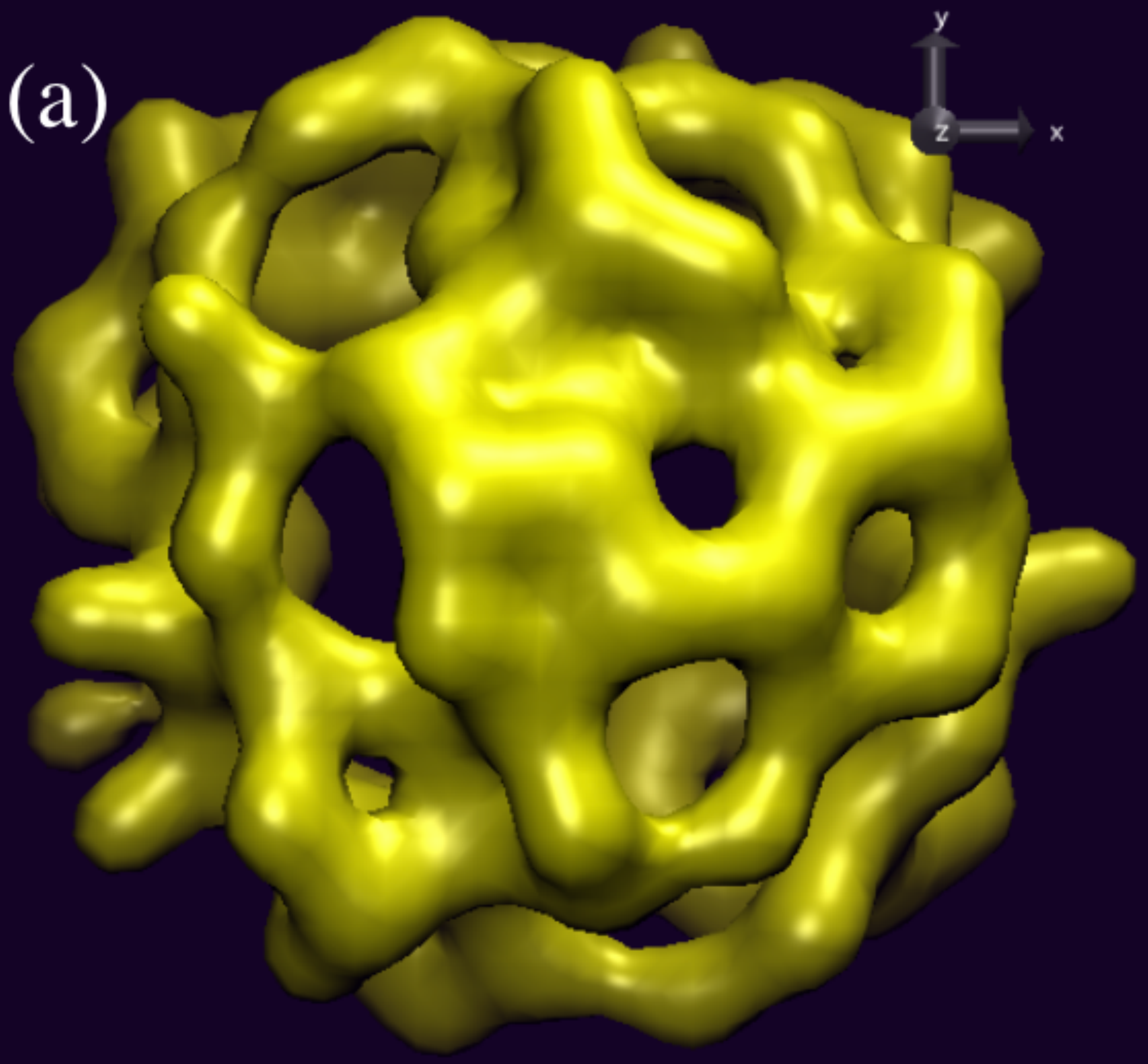}
\includegraphics[width= 0.4\textwidth]{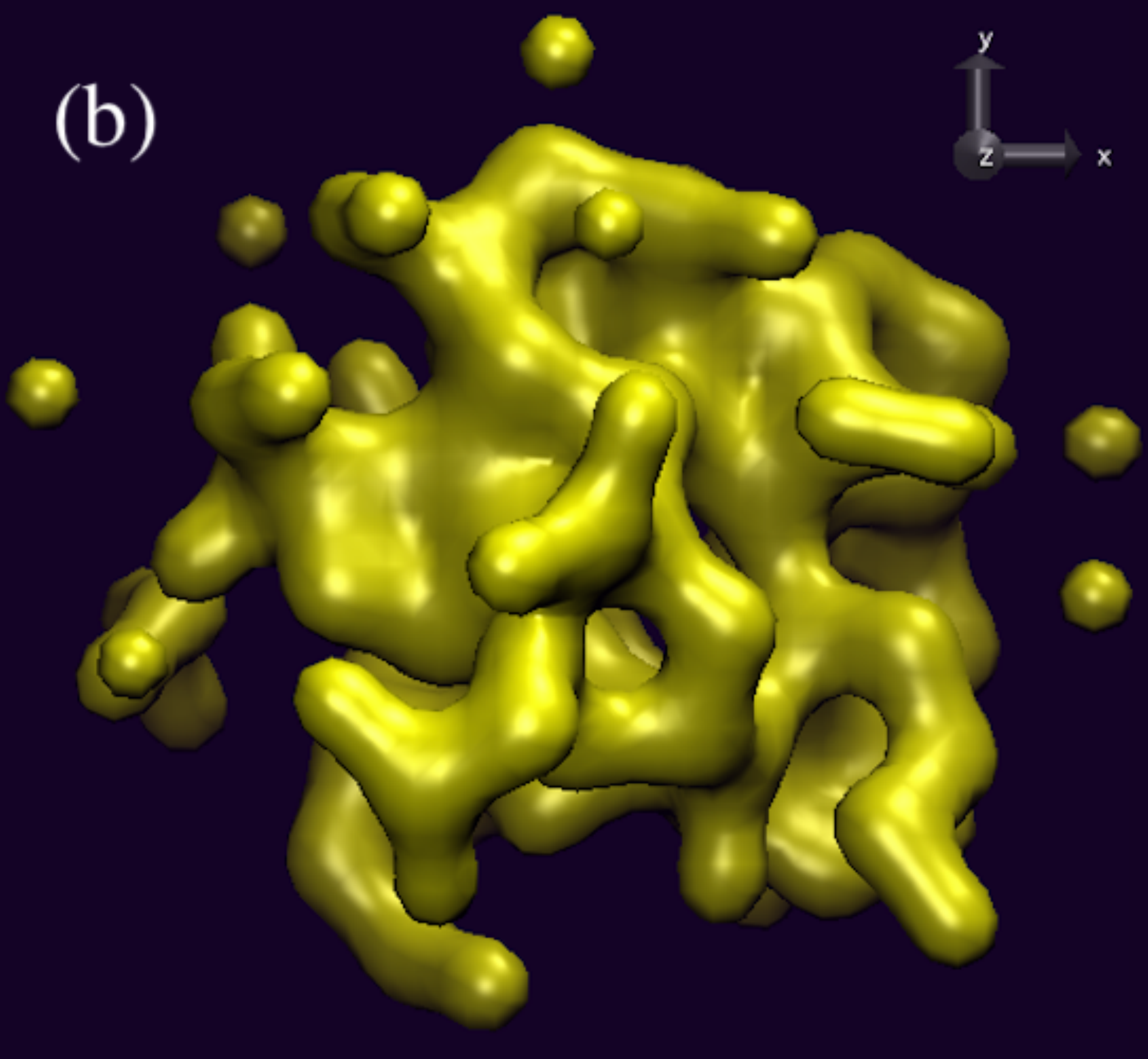}
\includegraphics[width= 0.38\textwidth]{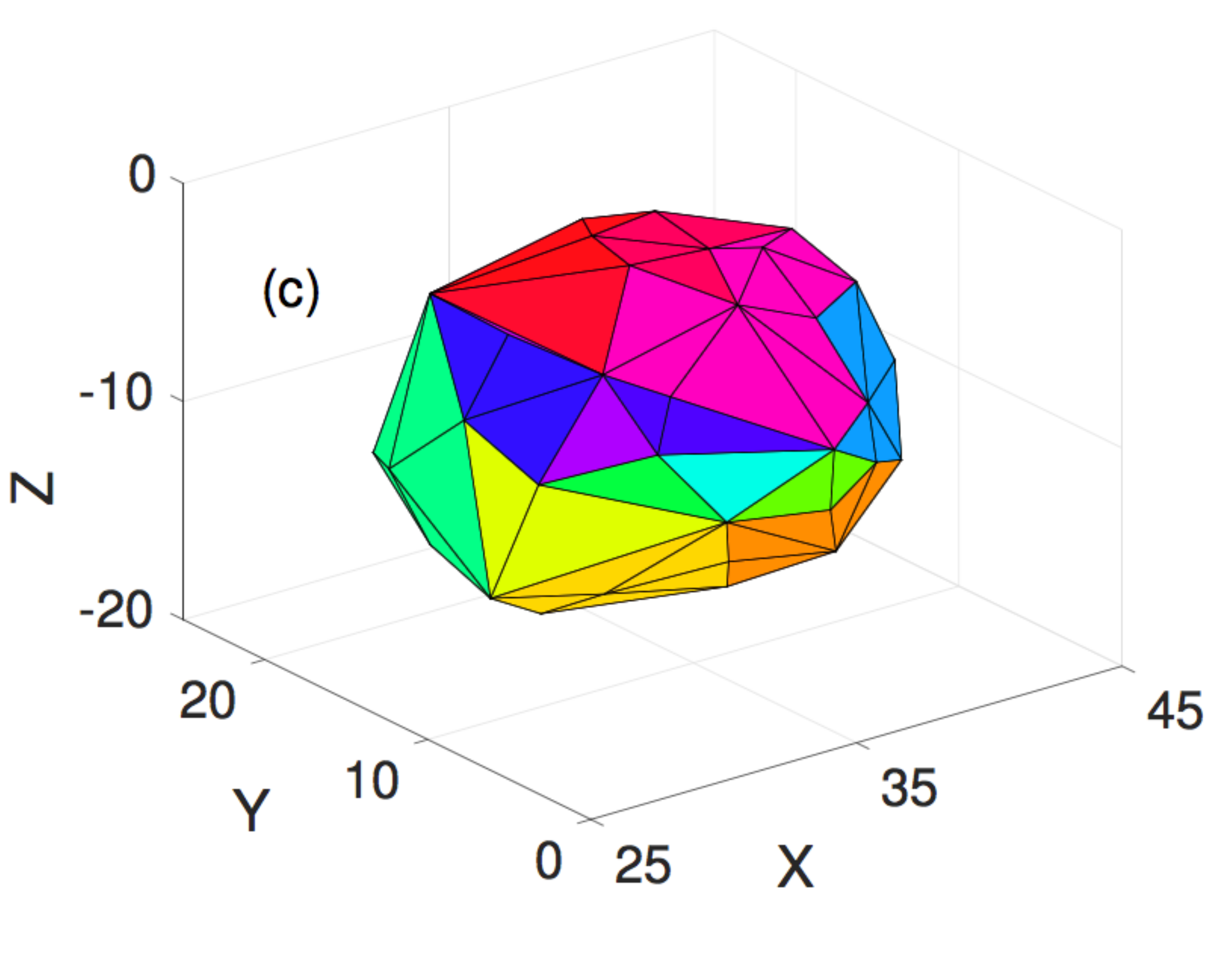}
\includegraphics[width= 0.38\textwidth]{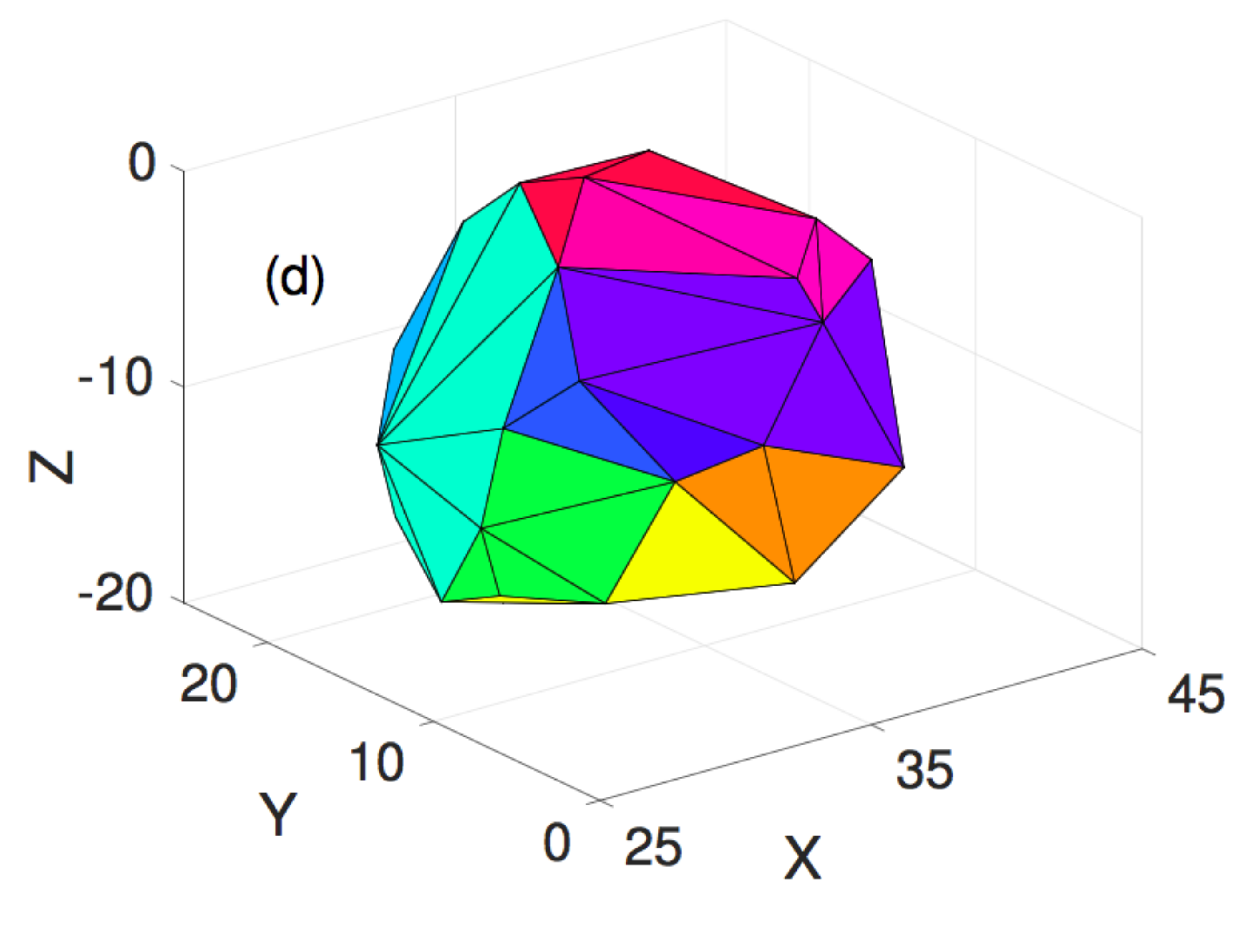}
\caption{
An isosurface representation of a void (V3) obtained 
from: (a) classical; and (b) {\it ab initio} simulations 
upon annealing at 800 K, followed by geometry relaxation. 
The corresponding shapes of the voids in the convex-hull 
approximation are shown in (c) and (d), respectively.
}
\label{FIG7}
\end{figure}

\begin{figure}[htb!]
\begin{center} 
\includegraphics[width=0.5\textwidth]{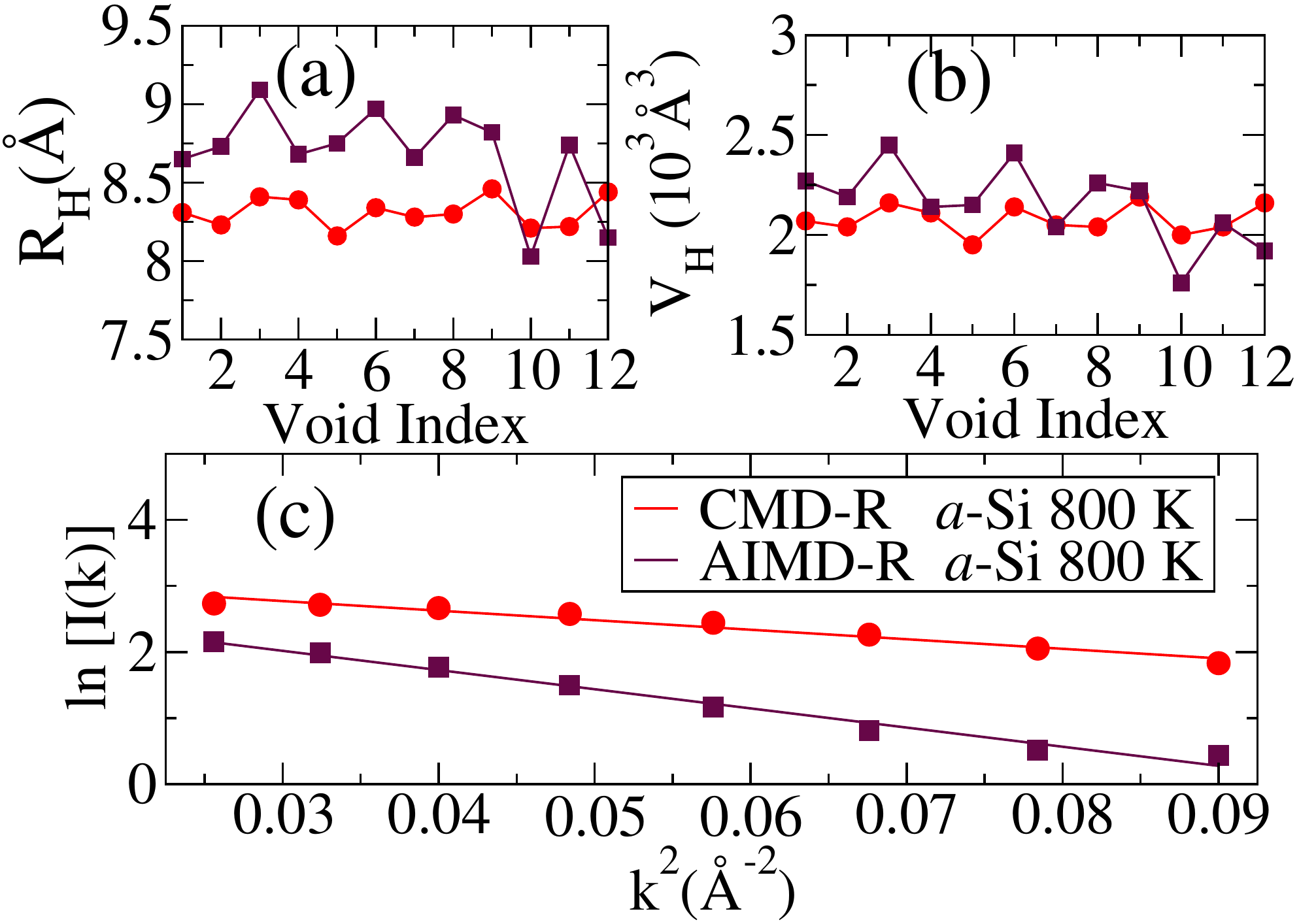}
\caption{
(a) The convex-hull radius for the voids from classical 
CMD-R (\textcolor{red}{$\bullet$}) and AIMD-R 
({\textcolor{maroon}{\small $\blacksquare$}}) simulations 
at 800 K, followed by geometry relaxation. (b) The corresponding 
volume of voids from the convex-hull approximation. 
(c) Guinier plots for the models at 800 K. The Guinier 
radii from the plots correspond to a value of about 
6.57 {\AA} (CMD-R) and 9.48 {\AA} (AIMD-R). 
} 
\label{FIG8}
\end{center} 
\end{figure}

Having studied the microstructure of voids at 300 K, we 
now examine the structure of voids at the high temperature 
of 800 K. Figure \ref{FIG7} shows the shape of a representative void 
(V3) after annealing at 800 K from classical and {\it ab initio} 
MD simulations. A comparison between the two structures in 
Figs.\,\ref{FIG7}(a) and \ref{FIG7}(b) shows the regions on 
the two surfaces with a varying degree of reconstruction. The 
shape of the respective void surfaces obtained from using the 
convex-hull approximation is also shown in Figs.\,\ref{FIG7}(c)-(d). 
Table \ref{TAB2} lists some of the parameters that characterize 
the void geometry in terms of the convex-hull radius ($R_H$), 
the radius of gyration of the void-surface atoms ($R_g$), the 
Guinier radius ($R_G$), the hull volume ($V_H$), and the 
average sphericity of the voids ($\langle \Phi_S \rangle$) 
after annealing at 300 K and 800 K. Although these scalar 
parameters do not vary much in the convex approximation, it 
is evident from the isosurface representation of the void in 
Figs.\,\ref{FIG7}(a) and \ref{FIG7}(b) that classical results 
noticeably differ from {\it ab initio} results, indicating 
the limitation of the classical force field in describing 
the restructuring of void surfaces at high temperature, namely 
at 800 K. This observation is also reflected in Figs.\,\ref{FIG8}a-c, 
where the convex-hull radii, the hull volumes, and the 
intensity of SAXS at 800 K, respectively, are presented.

\begin{table}[ht]
\caption{\label{TAB2} 
Estimated  volumes ({\AA}$^3$), linear sizes (\AA), 
and the average sphericities of the voids in {\asi} 
with and without hydrogen inside voids. $R_{X}$ 
indicates the radius of the void obtained from 
the convex-hull (X = H) and Guinier approximations (X = G), 
and the gyrational radius of the 
void-surface atoms (X = g).
} 
\begin{ruledtabular}
\begin{tabular}{cccccc}
Models & $R_{g}$ & $R_H$ & $R_G$ & $V_H$ & $\langle\Phi_S\rangle$  \\
\hline
\multicolumn{6}{c}{{\asi} (without hydrogen in voids)} \\
\hline
\hline
CMD-R300  & 7.60 & 8.34 & 6.38 & 2100 & 0.66\\
AIMD-R300 & 7.55 & 8.28 & 8.07 & 2040 & 0.67\\
CMD-R800  & 7.56 & 8.31 & 6.57 & 2080 & 0.63\\
AIMD-R800 & 6.84 & 8.67 & 9.48 & 2250 & 0.68\\
\hline
\multicolumn{6}{c}{{\asi} (with hydrogen in voids)} \\
\hline
\hline
AIMD-R300 & 6.66 & 8.36 & 8.66  & 2120 & 0.56\\
AIMD-R800 & 6.61 & 8.76 & 10.01 & 2290 & 0.71\\
\end{tabular}
\end{ruledtabular}
\end{table} 

\subsection{Structural evolution of hydrogen-rich voids in {\asi}}

Small-angle X-ray scattering measurements on HWCVD films 
indicate that the presence of preexisting H clusters can 
affect the nanostructure of voids due to H$_2$ bubble 
pressure in the cavities, which, upon annealing, can 
lead to geometric changes, as observed 
via tilting SAXS and surface transmission electron 
microscopy (STEM).\cite{Young2007} Although the degree 
of such nano-structural changes may depend on the growth 
mechanism and the rate of film growth, it is instructive 
to address the problem from a computational viewpoint 
using CRN models of {\asi} with nanometer-size voids, 
which are characterized by an experimentally 
consistent void-volume fraction and clusters of H 
atoms inside the voids. 

\begin{figure}[t!] 
\centering
\includegraphics[height=2.0 in, width=2.0 in]{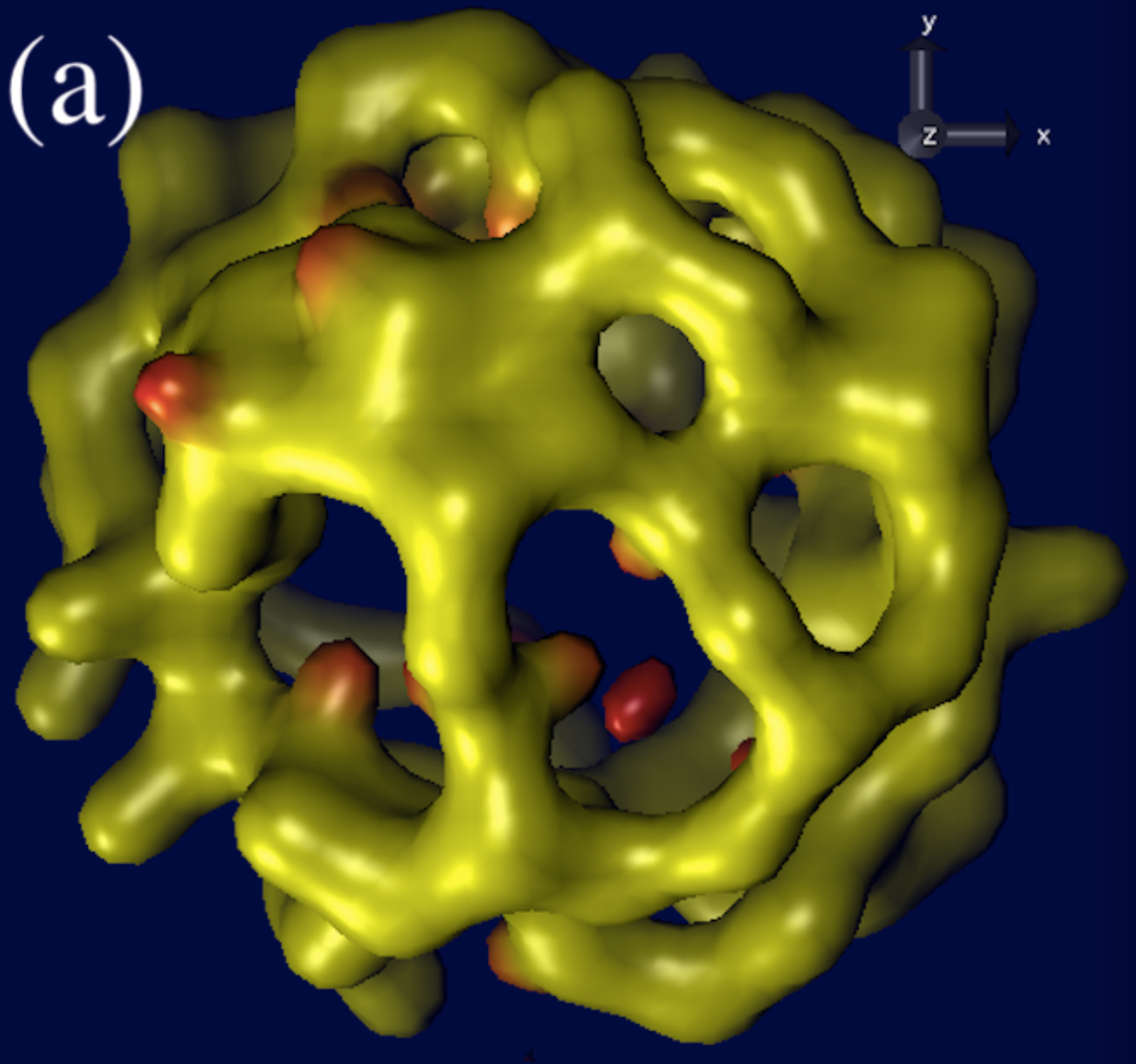}
\includegraphics[height=2.0 in, width=2.0 in]{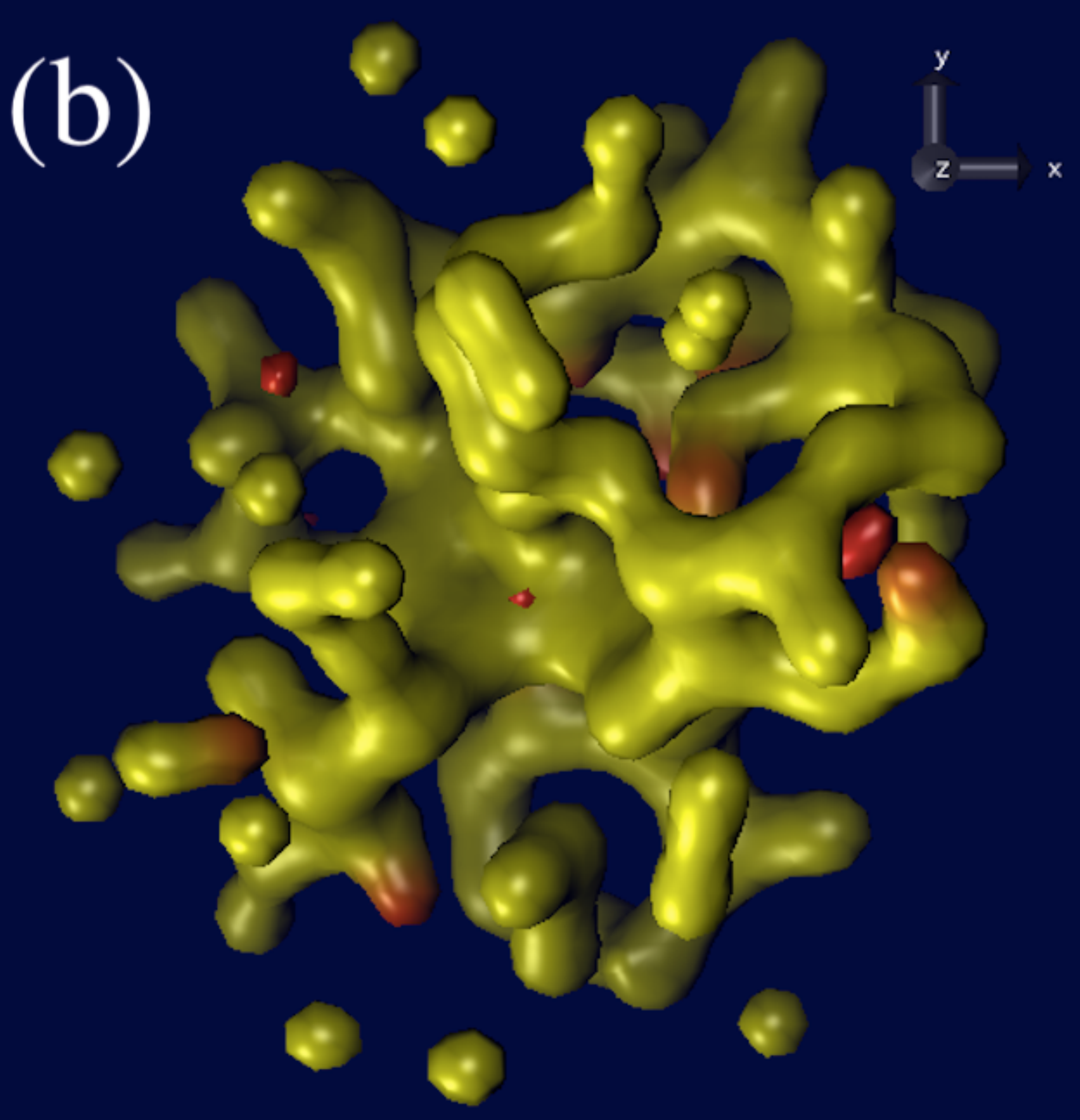}
\caption{
The structure of a representative void (V3), in isosurface 
representation, from AIMD simulations in the presence of 
H atoms inside the cavity at: (a) 300 K; and (b) 800 K. 
The yellow and red blobs represent Si and H atoms, respectively. 
For visualization and comparison on the same footing, an 
identical set of isosurface parameters was used to generate 
the surfaces. 
} 
\label{FIG9}
\end{figure}

Figure \ref{FIG9} shows the structure of a hydrogen-rich nanovoid 
surface (of V3) obtained after annealing at 300 K and 800 K, 
followed by total-energy minimization. The evolution of Si and 
H atoms during annealing was described using {\it ab initio} 
density-functional forces and energies, from the local-basis 
DFT package {\sc Siesta}.~\cite{siesta1} The procedure to conduct 
such AIMD simulations of voids, which are embedded in an ultra-large 
model of {\asi}, has been described in sec.\,IIC. 
A comparison of Fig.\,\ref{FIG9}(a) (with H atoms inside 
the void) with Fig.\,\ref{FIG4}(b) (with no H atoms) 
reveals that the thermalization of the void (V3) and its 
neighboring region at 300 K does not introduce notable 
changes on the void surface, with the exception of the 
distribution of H atoms.  The initial random distribution of H 
atoms, which was confined within a radius of 4 {\AA} from the 
void center, evolved to produce several monohydrides (SiH) and 
a few dihydrides (SiH$_2$), along with a few H$_2$ molecules 
inside the cavity. The majority of H atoms continue to stay near 
the void at 300 K.  By contrast, a considerable number of Si 
and H atoms have been found to be thermally driven out of 
the void region V3 at 800 K. An analysis of the distribution 
of Si atoms on the void surface, shown as yellow blob in 
Fig.\,\ref{FIG9}, and H atoms (red) inside the void  reveals 
that approximately 12.4\% of the original Si atoms and 
23.3\% of the H atoms left the void region of radius 10.5 {\AA} 
after annealing at 800 K.  This is apparent from the isosurface 
plot in Fig.\,\ref{FIG9}(b), where a considerable number 
of Si atoms can be seen to move away from the void-surface 
region, creating a somewhat diffused or scattered isosurface. 
This observation is found to be true for most of the other 
voids. On average, over 12 independent voids, approximately 
8.3\% of the Si atoms and 16.1\% of the 
H atoms were found to leave the void regions at 800 K.  
By contrast, the corresponding average values for the same 
at 300 K for Si and H atoms are 0\% and 2.6\%, respectively.  
It thus follows that at 800 K the restructuring of a 
void surface can be considerably affected via thermal 
and H-induced motion of Si atoms, as well as through the 
formation of various silicon-hydrogen bonding configurations on 
void surfaces. Table \ref{TAB3} presents the statistics 
of various silicon-hydrogen bonding configurations in 
the vicinity of each void, which is defined by a radius 
of 10.5 {\AA}.

\begin{table}[t]
\caption{\label{TAB3} 
Statistics of various silicon-hydrogen bonding configurations 
near hydrogen-rich voids at 300 K and 800 K from AIMD 
simulations. $R_H$ and $\Phi_S$ indicate the convex-hull 
radius and sphericity of the voids, whereas asterisked 
entries indicate at least one H atom left the void region 
of radial size 10.5 {\AA}.
}
\begin{ruledtabular}
\begin{tabular}{ccccccccccc}
Void&H$_{iso}$&H$_{2}$& SiH& SiH$_{2}$& SiH$_{3}$& $R_{H}$& $\Phi_S$  \\
\multicolumn{3}{c}{} \\
\hline
         &&&&   300 K \\
\hline
V1$^*$&2&1&20&2&0&8.34& 0.51\\
V2&0&3&22&1&0&8.33&0.56\\
V3&0&3&17&3&0&8.5&0.59\\
V4&0&0&26&1&0&8.4&0.58\\
V5&0&1&25&1&0&8.2&0.62\\
V6$^*$&1&2&22&1&0&8.34&0.54\\
V7&0&1&26&1&0&8.31&0.54\\
V8&1&1&23&2&0&8.5&0.57\\
V9&0&3&20&0&1&8.35&0.51\\
V10$^*$&1&3&18&2&0&8.28&0.60\\
V11&2&4&19&0&0&8.34&0.59\\
V12&2&2&23&0&0&8.52&0.56\\
\hline
   &&&&            800 K \\
\hline
V1&0&0&20&2&0&8.36&0.90\\
V2&0&2&20&2&0&8.84&0.66\\
V3$^*$&1&2&16&1&0&9.07&0.79\\
V4&0&1&23&1&0&8.64&0.62\\
V5&0&0&20&2&0&8.95&0.70\\
V6&0&2&14&1&1&8.67&0.83\\
V7&0&3&14&3&0&8.94&0.64\\
V8$^*$&1&1&14&3&0&8.95&0.73\\
V9&0&1&22&1&0&9.07&0.62\\
V10&0&1&21&2&0&8.71&0.62\\
V11&2&3&15&1&0&8.38&0.81\\
V12&1&0&21&2&0&8.56&0.56\\
\end{tabular}
\end{ruledtabular}
\end{table} 

\begin{table}[ht]
\caption{
The average concentration (in \% of H atoms) of various 
silicon-hydrogen bonding, H$_2$ molecules, and isolated H near 
voids (within a radius of 10.5 {\AA}). Bonded and non-bonded 
hydrogen that left the voids during annealing are listed under 
the category Ex-void.
\label{TAB4} 
}
\begin{ruledtabular}
\begin{tabular}{ccccccccccc}
Temp. &H$_{iso}$&H$_{2}$& SiH& SiH$_{2}$ &SiH$_{3}$ & Ex-void \\
\multicolumn{3}{c}{} \\
\hline
300 K & 2.5 & 13.3 & 72.5 & 8.3  & 0.8 & 2.6 \\
800 K & 1.4 & 8.9  & 61.1 & 11.7 & 0.8 & 16.1 \\
\end{tabular}
\end{ruledtabular}
\end{table} 
\begin{figure}[t!] 
\begin{center} 
\includegraphics[width=0.5\textwidth]{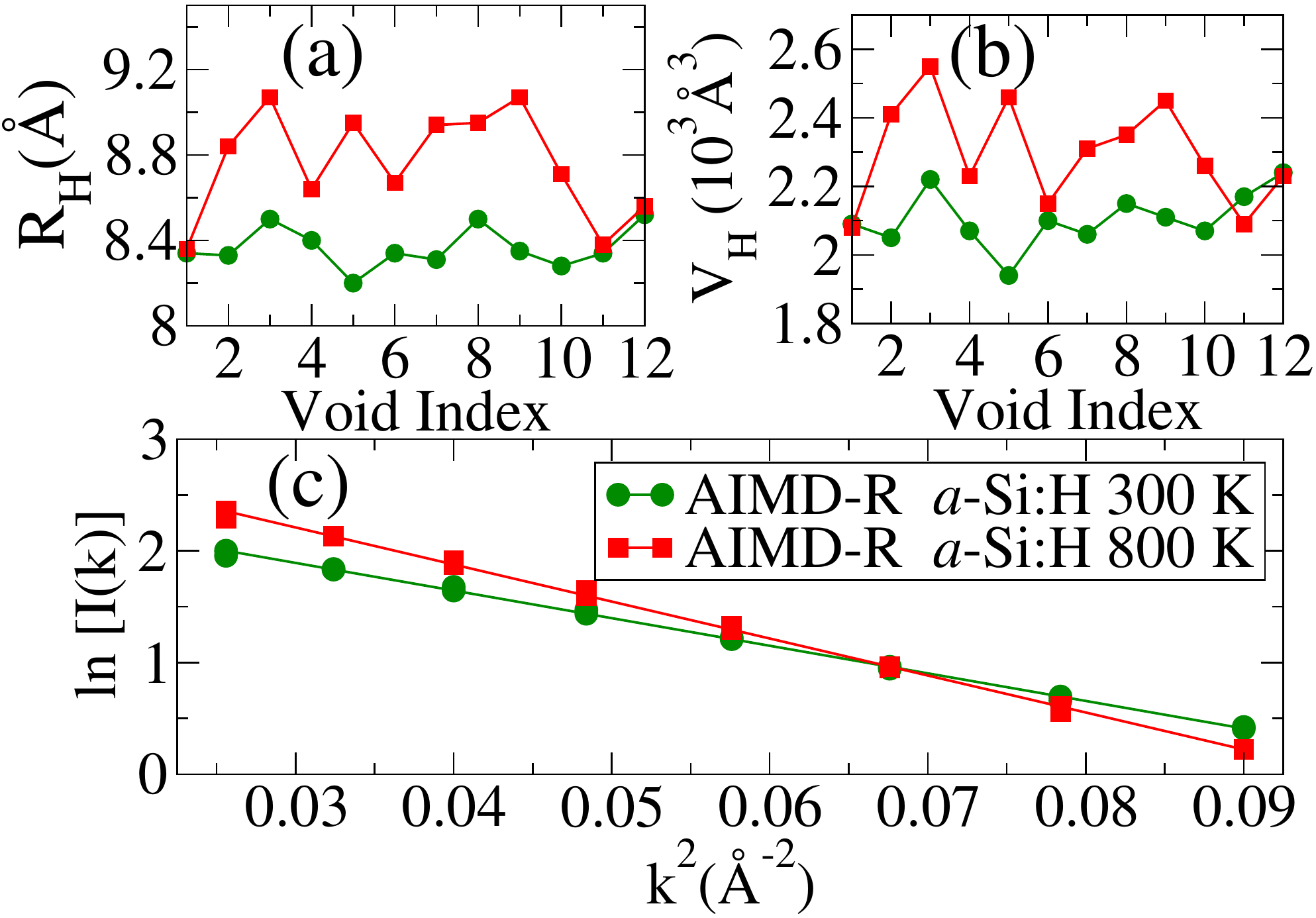}
\caption{
(a) The radius of gyration of hydrogen-rich voids from 
AIMD simulations at 300 K (\textcolor{green1}{$\bullet$}) 
and 800 K (\textcolor{red}{\small $\blacksquare$}) in the 
convex-hull approximation. (b) Estimated convex-hull volumes 
at 300 K and 800 K. (c) Guinier fits for the SAXS 
intensities in AIMD-R models of {\asih}  at 300 K 
and 800 K. 
} 
\label{FIG10}
\end{center} 
\end{figure}
To obtain the linear size of the voids, we computed the 
radius of gyration, $R_g$, using Eq.\,(\ref{4a}). 
Likewise, the convex-hull approximation of a void region, defined 
by a set of void-surface atoms, provides the radius of the convex 
polyhedron, $R_H$, from Eq.\,(\ref{4b}). 
Figure \ref{FIG10}(a)-(b) shows the values of $R_H$ and the 
corresponding hull volumes, $V_H$, of 12 voids, each 
of which has been computed upon annealing and geometry relaxation. 
Similarly, the $R_g$ values obtained from the real-space 
distribution of the void atoms are listed in Table \ref{TAB2}. 
The apparent deviation of $R_H$ from $R_g$ can be 
attributed to the fact that the computation of $R_g$ involves 
all the atoms on the surface and interior regions of the voids, 
whereas the convex-hull approximation includes only those atoms 
on the void-surface region that define the minimal 
convex polyhedral volume. 
Thus, $R_H$ provides an upper bound of the radial size of 
the voids. In order to compare $R_H$ and $R_g$ values with the 
linear size of the voids from the simulated intensity plots in 
SAXS, we have invoked the Guinier approximation to estimate the Guinier radius, 
$R_G$, from $\ln I(k)$-$k^2$ plots, as shown in Fig.\,\ref{FIG10}(c). 
Table \ref{TAB2} suggests that the $R_G$ values obtained from 
the simulated intensity plots reasonably match with the corresponding 
average values of $R_H$ at 300 K and 800 K. It may be noted that, 
while the intensity plot in Fig.\,\ref{FIG10}(c) is not 
particularly sensitive to the shape of the voids, the slightly 
pronounced scattering in the region $k^2 \le$ 0.05 {\AA}$^{-2}$ 
is possibly indicative of the expansion of the void volume at 800 K, 
as observed in the convex-hull volume in Fig.\,\ref{FIG10}(b). The 
exact cause of this increased scattering intensity is difficult 
to determine but it appears that a combination of the pressure 
due to H$_2$ bubbles, surface roughening of the voids, and the 
displacement of the atoms on the void surfaces at high temperature 
could lead to this additional scattering.  This observation is 
corroborated by the experimental results from a small-angle 
X-ray scattering study of nanovoids in HWCVD amorphous Si films 
that indicated an increase of individual void volumes upon 
annealing at 813 K.~\cite{Young2007}

\subsection{Hydrogen dynamics near voids in {\asi}}

In this section, we shall make a few observations 
on the dynamics of hydrogen atoms in the 
vicinity of void surfaces from {\it ab initio} 
molecular-dynamics simulations.  Since the mass of H atoms is 
significantly smaller than that of Si atoms, the motion of H 
atoms is more susceptible 
to the annealing temperature and, more importantly, to 
the resulting temperature-induced structural changes on the void 
surfaces. This is particularly noticeable at 
high temperature.
Having established earlier that the structural changes 
on the void surfaces are more pronounced at 800 K, 
we shall now examine the character of hydrogen 
dynamics inside (and near the surface of) the voids 
at 800 K. Despite the limited timescale of 
the present AIMD simulations, spanning 10 ps 
only, it is reasonable to assume that, at high 
temperature, the dynamical character of hydrogen motion 
inside the voids would be reflected in the first 
several picoseconds.  It may be noted that, unlike 
structural properties, the dynamics of hydrogen in 
{\asi} may depend on the accuracy of total energies 
and forces from DFT calculations. The use of the 
Harris-functional approximation in the present 
work, in an effort to address the structure of 
voids on the nanometer length scale, suggests 
that the results on hydrogen dynamics presented 
here are somewhat approximate in nature. A full 
self-consistent-field DFT calculation, over a 
period of a few tens of picoseconds, using plane-wave 
or even local basis sets for 2200-atom models 
is computationally overkill and outside scope of 
the present work on nanostructural evolution. 

\begin{figure*}[t!]
\begin{center}
\includegraphics[width = 0.42\textwidth]{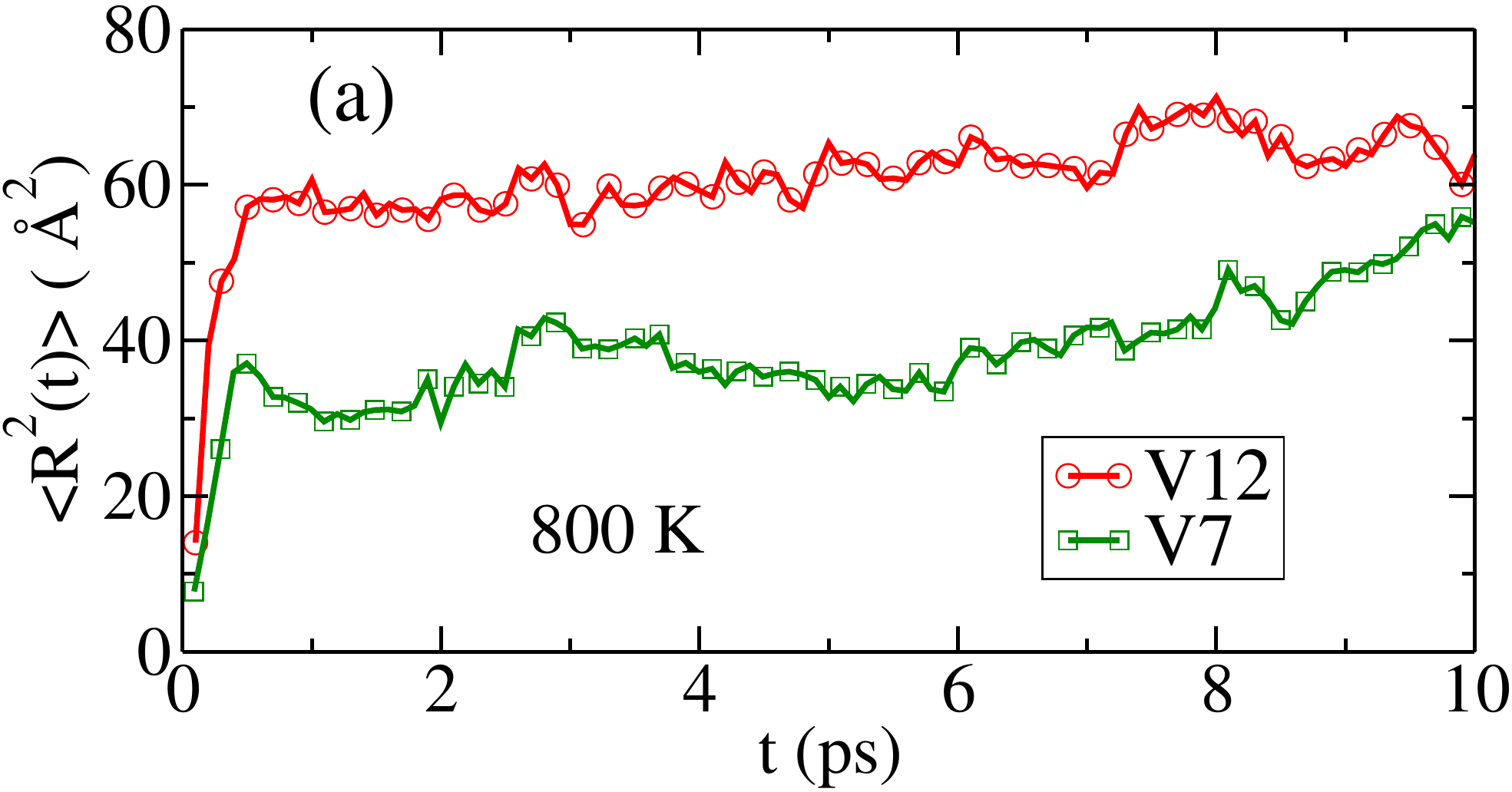}
\includegraphics[width = 0.25\textwidth]{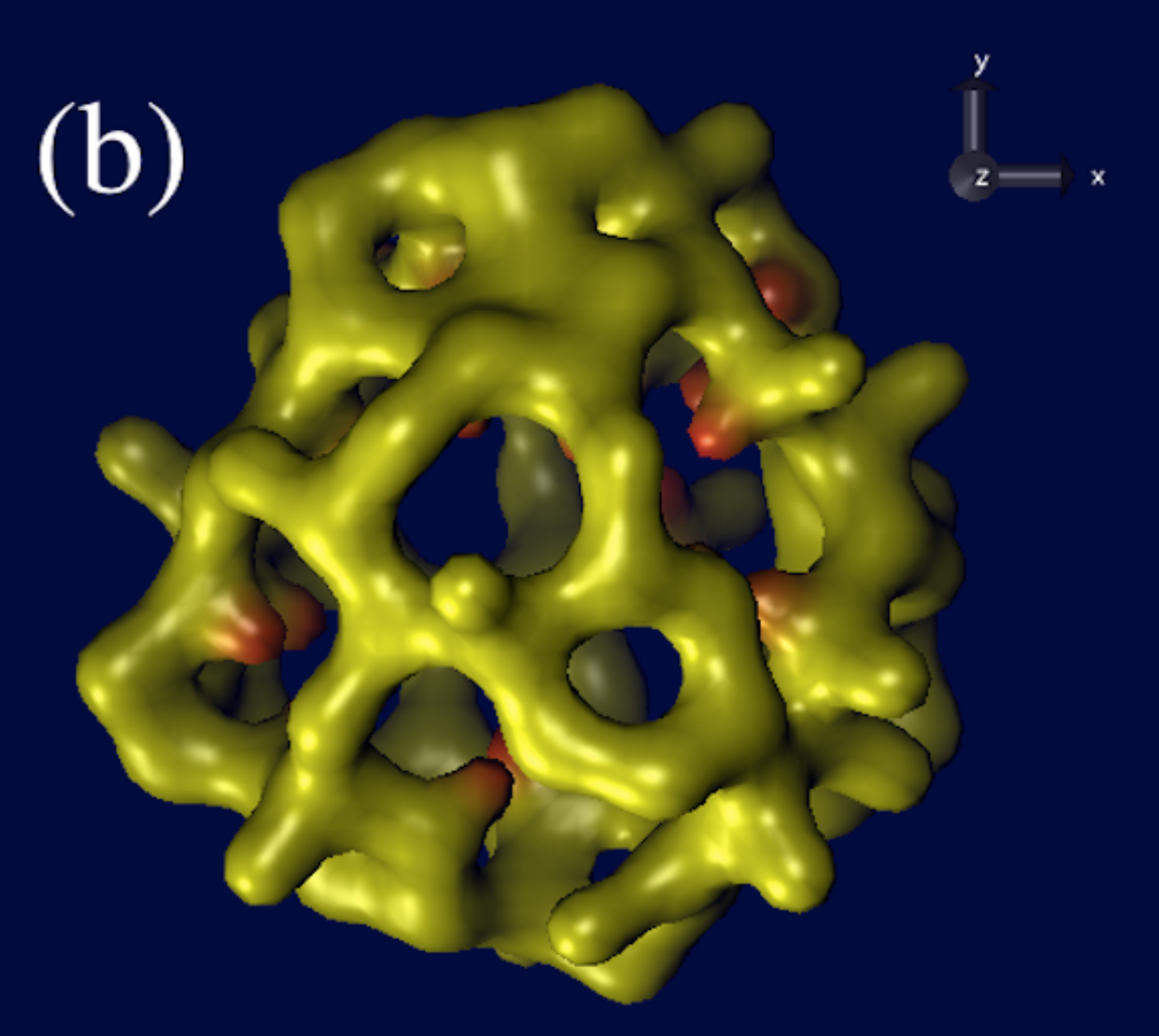}
\includegraphics[width = 0.253\textwidth]{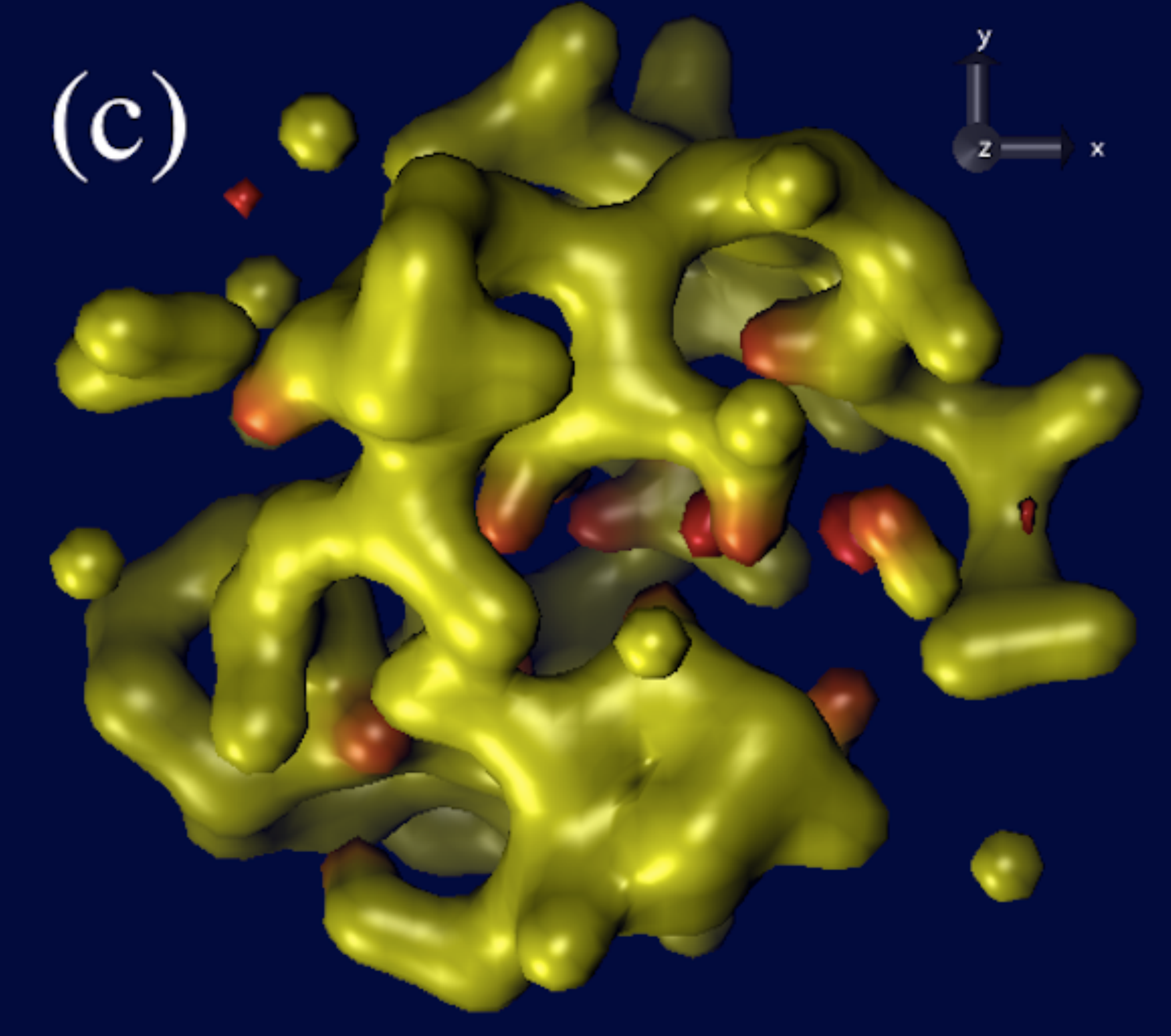}
\caption{
(a) The time evolution of the mean-square displacements 
(MSD) of H inside two voids, V12 (\textcolor{red}{$\circ$}) 
and V7 (\textcolor{green1}{$\Box$}), at 800 K. The 
difference between the two sets of MSD values, from 1 ps 
to 8 ps, can be attributed to the degree of void-surface 
restructuring, as discussed in sec. IIIC. 
(b) The compact or smooth structure of the 
V12 surface, obtained from an isosurface representation 
of the void, compared to a relatively diffused or 
scattered distribution of Si atoms in (c) V7 at 800 K.
}
\label{H2}
\end{center}
\end{figure*}

Figure \ref{H2}(a) depicts the time evolution of 
hydrogen atoms at 800 K within two voids, V12 and 
V7. Initially, all hydrogen atoms/molecules 
were uniformly 
distributed inside the voids within a radius of 
4 {\AA} from the center of the voids (see 
Fig.\,\ref{void}).  The mean-square displacements 
(MSD), $\langle R^2(t)\rangle$, of hydrogen 
atoms -- averaged over all H atoms in the respective 
void -- are shown in Fig.\,\ref{H2}(a) from 0 to 
10 ps. The corresponding real-space distributions 
of Si and H atoms on the void surfaces and their 
immediate vicinity are also shown in 
Figs.\,\ref{H2}(b) and \ref{H2}(c) for V12 and 
V7, respectively. The early-time behavior of the 
dynamics, from 0 to 0.5 ps, can be roughly understood 
from elementary considerations. Given that the majority of H atoms 
are away from the void surface at the beginning 
of simulation, by at least 2 {\AA} or more, the 
early-time behavior of the dynamics would be 
primarily determined by the temperature 
of the system for a reasonably well-defined initial H 
distribution. Since the root-mean-square (RMS) speed of H 
atoms at 800 K for the Maxwell-Boltzmann distribution is about 
0.045 {\AA}/fs, it would be reasonable to assume that 
the initial behavior can last for about a fraction of a picosecond, 
after which H atoms become affected by the presence of 
the void surface. 
It is the marked difference in the value 
of $\langle R^2(t)\rangle$ in Fig.\,\ref{H2}, from 1 ps to 8 ps, 
that we find most interesting.  This notable difference between the 
two sets of MSD values beyond 0.5 ps can be understood 
by taking into account the real-space structure of the 
void surfaces. In addition to the trivial thermal and configurational 
fluctuations -- caused by temperature and the disorder 
in H distributions, respectively -- the motion of H atoms 
is also affected by the varying degree of re-structuring 
of Si atoms on the void surfaces after 0.5 ps. 
The surface structure of V7 can be seen to be more 
open or scattered compared to its counterpart in 
V12, which is more compact or well-defined, as shown 
in Figs.\,\ref{H2}(c) and \ref{H2}(b), respectively.  
This has been also confirmed 
by analyzing the distributions of void-surface atoms in 
the interior of V12 and V7. The presence of Si atoms inside 
V7 may reduce the effective 
diffusion rate of H atoms, due to additional scattering 
from interior Si atoms, during intermediate-time evolution 
from 1 ps onward.  By contrast, the more well-defined 
void surface of V12, with few Si atoms inside, 
provides less resistance to H atoms during their intermediate-time 
evolution.  Thus, hydrogen diffusion in V12 proceeds 
relatively uninterruptedly, in the presence of few Si atoms from 
void surfaces, leading to a notable difference in their 
MSD values during the first several picoseconds of simulations. 
This observation has been found to be true for other pairs 
of voids for which a notable difference in the MSD 
values exists. 
In view of our observation that the intermediate-time 
behavior (i.e., from 1 ps to 8 ps) of the H dynamics 
can be influenced by the roughness of void surfaces, 
it might be necessary to take into account the appropriate 
size and the accurate surface structure of the voids in 
the calculations. However, a full SCF study of hydrogen 
dynamics inside {\it nanometer-size} voids in {\asi} 
for several tens of picoseconds is computationally 
prohibitive and outside the scope of the present study.

\subsection{Silicon-hydrogen bonding configurations on void surfaces}
Experimental data from positron-annihilation lifetime (PAL) 
spectroscopy,~\cite{Sekimoto2016} Rutherford backscattering 
spectrometry (RBS),~\cite{Sekimoto2018} and Fourier-transform 
infrared spectroscopy-attenuated total reflection 
(FTIR-ATR)~\cite{Sekimoto2018} indicate that bonded and non-bonded 
hydrogens play an important role in characterizing the 
structural and optical properties of {\asih}. Sekimoto 
{\etal}~\cite{Sekimoto2016,Sekimoto2018} recently demonstrated 
that the presence of non-bonded hydrogens (NBH) at concentrations 
beyond 2.8 at.\,\% in amorphous silicon networks led to changes 
in the vacancy-size distribution and induced formation of 
nanometer-size voids in the network via relaxation of internal 
stress.  The formation of H$_2$ molecules and their evolution 
can be probed by an analysis of the low-temperature (LT) peak 
of the hydrogen-effusion profile~\cite{Sekimoto2016,Beyer1991,Beyer2003} 
near 750 K, whereas the number density of H$_2$ molecules can 
be estimated from the collision-induced weak IR absorption in 
the frequency region of 4100 cm$^{-1}$ to 5500 cm$^{-1}$. Chabal 
and Patel~\cite{Chabal1984} reported a value of $10^{21}$ H$_2$/cm$^{3}$ 
by analyzing data obtained from IR measurements at room temperature.  
Although the present study, based on pure {\asi} networks, 
does not permit us to address the crucial role of silicon-hydrogen 
bonding configurations in {\asih} in general, it does provide 
us with new insights into the role of bonded and non-bonded 
hydrogens in the vicinity of the walls of H-rich voids. 
Tables \ref{TAB3} and \ref{TAB4} list the statistics and 
the average concentration of various bonded and non-bonded 
hydride configurations inside the voids, respectively. 
Non-bonded hydrogens chiefly comprise H$_2$ 
molecules, along with one or two isolated H atoms near the 
voids. The latter possibly originate from the short annealing 
time of 10 ps of AIMD simulations. The number of H$_{2}$ 
molecules per void ranges from 0 to 4 and 0 to 3 at 300 K 
and 800 K, respectively, which translates into a density 
of 1.5--2.2 $\times$ 10$^{21}$ H$_2$/cm$^3$ for nanometer-size 
voids.  This value matches very well with the experimental 
value of 10$^{21}$ H$_2$/cm$^3$ from IR measurements by 
Chabal and Patel,~\cite{Chabal1984} as mentioned earlier. 
\begin{figure}[t!] 
\begin{center} 
\includegraphics[height=2.2 in, width=2.2 in]{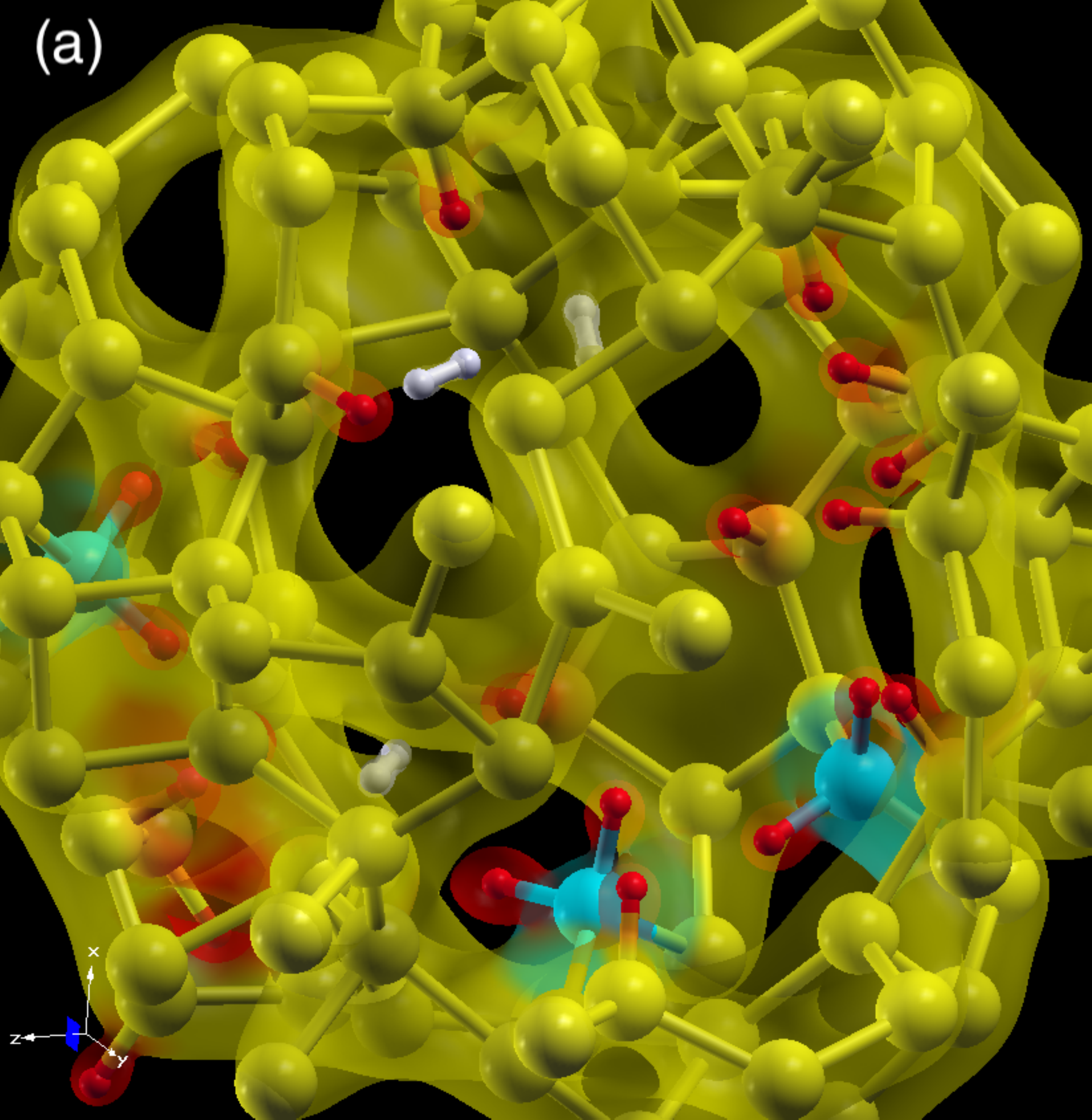} \hskip 0.2cm 
\includegraphics[height=2.2 in, width=2.2 in]{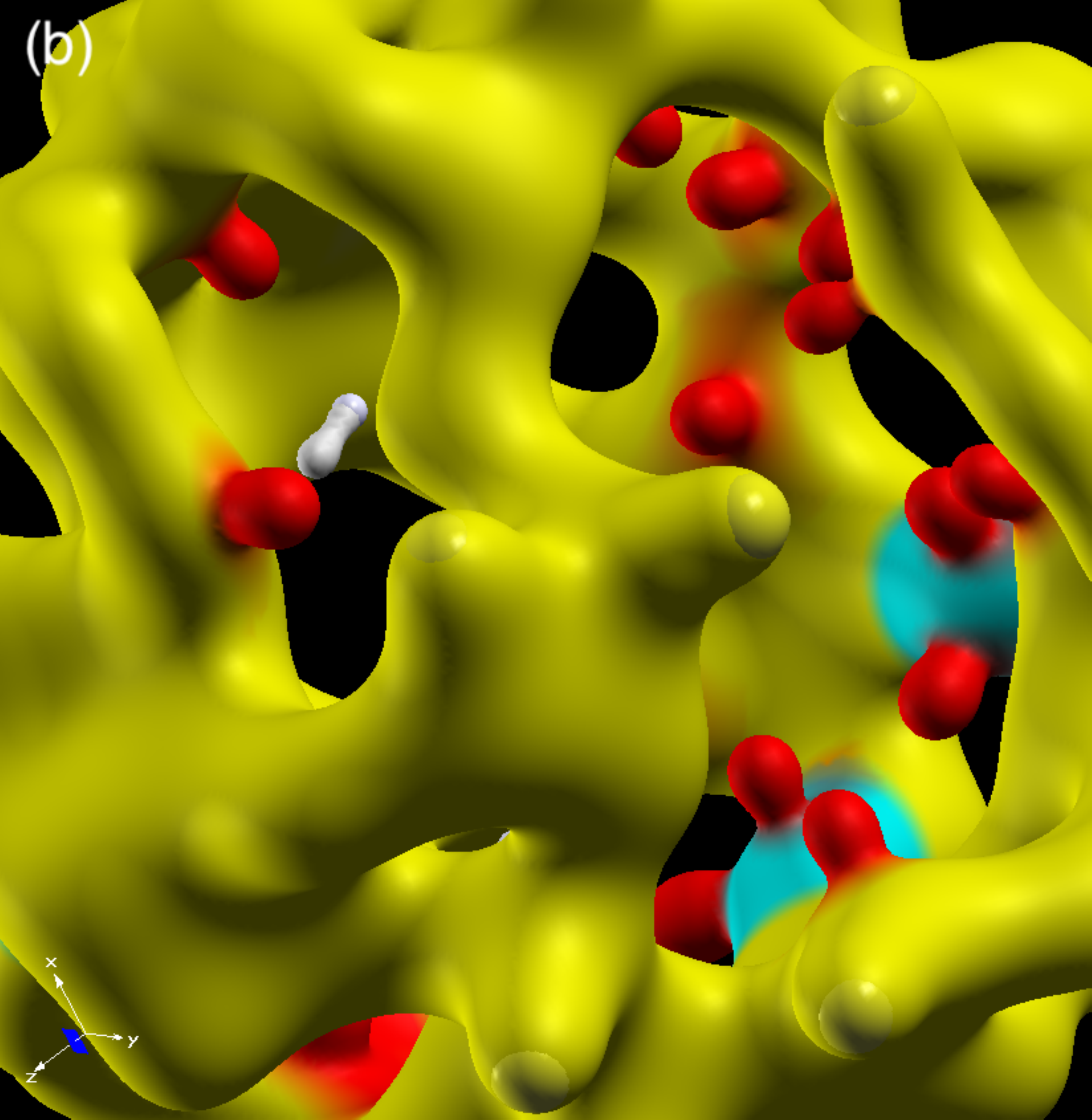}
\caption{
Bonded and non-bonded hydrogens inside a void (V3) from 
AIMD simulations at 300 K. (a) The interior wall of the void 
can be seen to be decorated with several SiH (yellow-red) 
and three SiH$_2$ (blue-red) configurations, as well as 
three H$_2$ molecules (white). (b) A magnified view, 
showing protruding monohydrides (yellow-red) 
and dihydrides (blue-red) on the walls, and a hydrogen
molecule (white) inside the void. 
}
\label{FIG12}
\end{center} 
\end{figure} 
These results are also consistent with the reported data 
from RBS measurements.~\cite{Sekimoto2016} 
\begin{figure}[t!] 
\centering
\includegraphics[height=2.0 in, width=3.0 in]{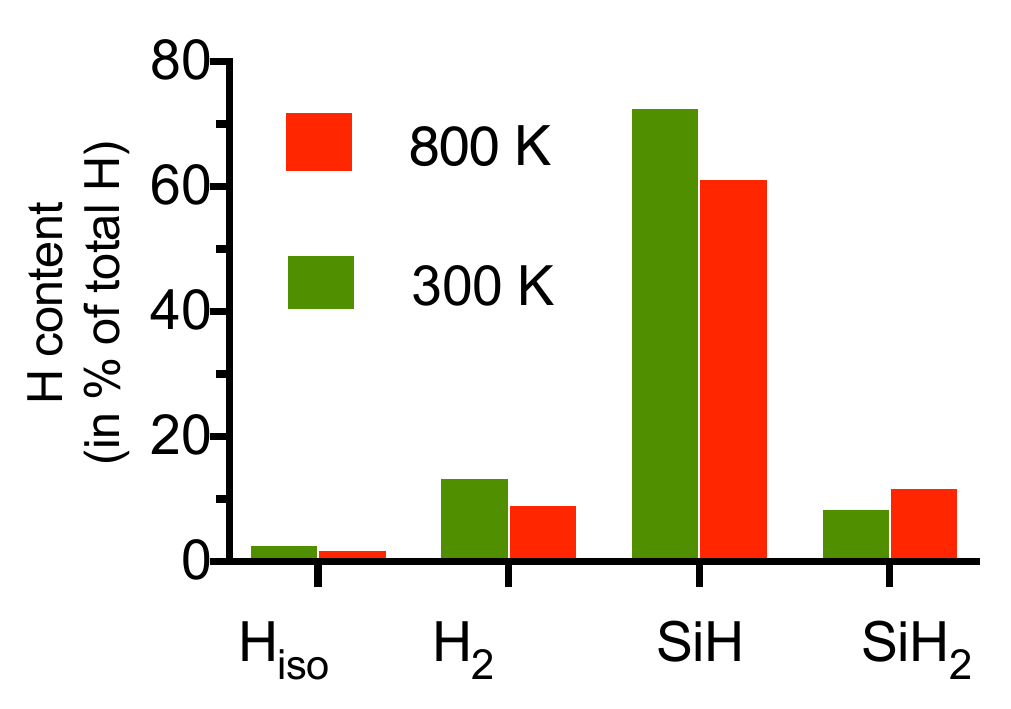}
\caption{
Average statistics of silicon-hydrogen bonding 
configurations near void surfaces in {\asih} 
after annealing at 300 K and 800 K. The results 
were obtained by averaging over 12 independent 
voids, distributed over a distance of several 
nanometers. 
}
\label{FIG13}
\end{figure} 
On the other hand, the great 
majority of bonded hydrogens appeared in monohydride (SiH) 
configurations. An examination of the local bonding environment 
of silicon atoms near the void surfaces showed that 
approximately 72.5 at.\,\% and 61.1 at.\,\% of total 
H atoms formed Si-H bonds at 300 K and 800 K, 
respectively (see Table \ref{TAB4}).  
Of the remaining H atoms, about 8.3 \% (11.7\%) were 
found to be bonded with Si as SiH$_2$ configurations 
at 300 K (800 K). Apart from a few isolated H atoms, the 
rest of the hydrogen has left the spherical void region of 
radius 10.5 {\AA}. 
A graphic distribution 
of the presence of H$_2$ molecules and the bonded 
SiH$_2$ configurations on the wall of V3 at 300 K is 
depicted in Fig.\,\ref{FIG12}. Several SiH$_2$ configurations 
can be seen to have formed on the surface of the void. This observation is 
consistent with experimental results from infrared measurements~\cite{Chabal1984} 
and the recent computational studies based on information-driven atomistic 
modeling of {\asih}.~\cite{PBiswas2015,BiswasJAP2014} 
Figure \ref{FIG13} shows the concentration of hydrogen 
associated with various silicon-hydrogen bonding configurations. 
The results clearly establish that a considerable 
number of H$_2$ molecules, of about 9--13\% of the 
total H, can form in the vicinity of voids, depending 
upon the annealing temperature, concentration of H 
atoms, the source of hydrogen, and the preparation 
methods of {\asi} samples. 

\section{Conclusions}
In this paper, we studied the temperature-induced 
nano-structural changes of the voids in {\asi} at 
low and high annealing temperature in the range 
300--800 K using classical and quantum-mechanical 
simulations, in the absence and presence of 
hydrogen near voids. 
This was achieved by generating an ultra-large 
model of {\asi}, consisting of more than 260,000 atoms, in 
order to be able to produce a realistic 
distribution of nanometer-size voids in the amorphous 
environment of Si atoms, as observed in SAXS, PAL, IR, 
and hydrogen-effusion measurements. 
An examination of the distribution of the atoms near the voids 
reveals that the  reconstruction of void surfaces in pure 
{\asi} at 300 K led to minimal changes in the shape and size 
of the voids, which can be readily understood from a 
classical treatment of the problem.
By contrast, the high-temperature annealing at 800 K caused 
significant changes in the shape and size of the voids, 
with noticeable structural differences between 
classical and quantum-mechanical results. This observation 
appears to indicate that classical potentials, such as 
the modified SW potential, might not be particularly 
suitable for an accurate description of the dynamics 
of Si atoms near voids and the resulting reconstruction 
of the void surfaces at high temperatures. 

An important outcome of the present study is that the dynamics of 
bonded and non-bonded hydrogens near the voids and the 
degree of void-surface reconstruction are found to be 
intrinsically related to each other. {\it Ab initio} 
annealing of {\asi} networks with a void-volume fraction of 
0.2\% at 300 K and 800 K suggests that the presence of hydrogen 
within voids can facilitate surface 
reconstruction through the formation of silicon monohydride 
and dihydride configurations, as well as the displacement 
of Si atoms on the void surfaces. The rearrangement 
of atoms on void surfaces affects the diffusion of 
non-bonded hydrogens, which in turn produce surface 
bumps and changes the shape and size of the original void. 
This observation is reflected in the time evolution of 
hydrogen during annealing. 
The presence of Si atoms inside a heavily 
reconstructed fuzzy/rough void surface reduces the effective 
diffusion rate of hydrogen, due to additional scattering from 
the interior Si atoms, in their intermediate-time evolution 
from 1 ps onward.  In contrast, a compact void surface 
provides less resistance to hydrogen 
during their evolution. Thus, hydrogen diffusion inside 
a smooth or well-defined void proceeds with little or no 
hindrance, leading to a notable difference in their 
mean-square-displacement (MSD) values during their 
intermediate-time evolution.  Finally, the results from 
the study show the presence of bonded hydrogens (BH), 
mostly SiH, SiH$_2$, and non-bonded hydrogens (NBH) 
in the form of H$_2$ molecules. 
The concentration of BH and NBH found in our study 
is consistent with the experimental values 
obtained from infrared (IR), Rutherford back 
scattering (RBS), and hydrogen-forward scattering (HFS) 
measurements.  An examination of the silicon-hydrogen bonding 
configurations near the voids suggests that the interior 
walls of the voids are decorated with SiH$_2$, which is 
supported by experimental results from infrared and 
ellipsometric studies. Likewise, the concentration of 
H$_2$ molecules obtained from the first-principles 
density-functional simulations in the present study is 
found to be consistent with the experimental value 
estimated from the collision-induced weak IR absorption 
by H$_2$ molecules in the frequency region of 4100--5500 cm$^{-1}$. 

\section*{acknowledgements}
This work was partially supported by the U.S. National Science
Foundation under Grants No.\;DMR 1507166 and No.\;DMR 1507118. 
We acknowledge the Texas Advanced Computing Center at the 
University of Texas at Austin for providing HPC resources 
that have contributed to the results reported in this work. 
One of us (P.B.) thanks Prof.\, David Drabold (Ohio University) 
for discussions. 

%

\end{document}